\documentclass[twocolumn,superscriptaddress]{revtex4-1}
\usepackage{color}
\usepackage{graphics,graphicx,epsfig}
\usepackage{epsf,epstopdf,wrapfig}
\usepackage{amssymb,amsfonts,amsmath}
\usepackage[export]{adjustbox}
\include{epsf}
\usepackage{ifthen}		

\newcommand{\beginsupplement}{%
        \setcounter{table}{0}
        \renewcommand{\thetable}{S\arabic{table}}%
        \setcounter{figure}{0}
        \renewcommand{\thefigure}{S\arabic{figure}}%
     }

\def\(({\left(}
\def\)){\right)}                       
\def\[[{\left[}
\def\]]{\right]}

\newcommand{\beqn}{\begin{eqnarray}}
\newcommand{\eeqn}{\end{eqnarray}}
\newcommand{\beq}{\begin{equation}}
\newcommand{\eeq}{\end{equation}}

\newcommand{\abs}[1]{|#1|}
\newcommand{\<}{\langle}
\renewcommand{\>}{\rangle}

\newcommand{\bs}{ \mbox{\boldmath$\sigma$}}

\renewcommand{\phi}{\varphi}

\definecolor{junglegreen}{rgb}{0.16, 0.67, 0.53}
\definecolor{myrtle}{rgb}{0.13, 0.26, 0.12}
\definecolor{lincolngreen}{rgb}{0.11, 0.35, 0.02}
\definecolor{forestgreen}{rgb}{0.13, 0.55, 0.13}

\newcommand{\presadd}{Present adress: School of Biological Sciences, Georgia Institute of Technology, Atlanta, GA, USA and Institut de Biologie, École Normale Supérieure, Paris, France}

\providecommand\JournalTitle[1]{#1}

\graphicspath{{fig/}}
\begin{document}

\title{From evolution to folding of repeat proteins}

\author{Ezequiel A. Galpern}
\affiliation{Protein Physiology Lab, Universidad de Buenos Aires, Facultad de Ciencias Exactas y Naturales, Departamento de Qu\'\i mica Biol\'ogica. Buenos Aires, Argentina. / CONICET - Universidad de Buenos Aires. Instituto de Qu\'\i mica Biol\'ogica de la Facultad de Ciencias Exactas y Naturales (IQUIBICEN). Buenos Aires, Argentina}
\author{Jacopo Marchi}
\affiliation{Laboratoire de physique de l'\'Ecole normale sup\'erieure
  (PSL University), CNRS, Sorbonne Universit\'e, and Universit\'e de
  Paris, 75005 Paris, France}
\affiliation{ \presadd}
\author{Thierry Mora}
\affiliation{Laboratoire de physique de l'\'Ecole normale sup\'erieure
  (PSL University), CNRS, Sorbonne Universit\'e, and Universit\'e de
  Paris, 75005 Paris, France}
 
\author{Aleksandra M. Walczak}
\affiliation{Laboratoire de physique de l'\'Ecole normale sup\'erieure (PSL University), CNRS, Sorbonne Universit\'e, and Universit\'e de Paris, 75005 Paris, France}

\author{Diego U. Ferreiro}
\affiliation{Protein Physiology Lab, Universidad de Buenos Aires, Facultad de Ciencias Exactas y Naturales, Departamento de Qu\'\i mica Biol\'ogica. Buenos Aires, Argentina. / CONICET - Universidad de Buenos Aires. Instituto de Qu\'\i mica Biol\'ogica de la Facultad de Ciencias Exactas y Naturales (IQUIBICEN). Buenos Aires, Argentina}

\begin{abstract}

Repeat proteins are made with tandem copies of similar amino acid stretches that fold into elongated architectures. Due to their symmetry, these proteins constitute excellent model systems to investigate how evolution relates to structure, folding and function. Here, we propose a scheme to map evolutionary information at the sequence level to a coarse-grained model for repeat-protein folding and use it to investigate the folding of thousands of repeat-proteins. We model the energetics by a combination of an inverse Potts model scheme with an explicit mechanistic model of duplications and deletions of repeats to calculate the evolutionary parameters of the system at single residue level. This is used to inform an Ising-like model that allows for the generation of folding curves, apparent domain emergence and occupation of intermediate states that are highly compatible with experimental data in specific case studies. We analyzed the folding of thousands of natural Ankyrin-repeat proteins and found that a multiplicity of folding mechanisms are possible. Fully cooperative all-or-none transition are obtained for arrays with enough sequence-similar elements and strong interactions between them, while non-cooperative element-by-element intermittent folding arose if the elements are dissimilar and the interactions between them are energetically weak. In between, we characterised nucleation-propagation and multi-domain folding mechanisms. Finally, we showed that stability and cooperativity of a repeat-array can be quantitatively predicted from a simple energy score, paving the way for guiding protein folding design with a co-evolutionary model.
\end{abstract}

\maketitle

\section{Introduction}
Robust folding and long-term evolution are two of the most basic aspects of natural protein molecules. These features are necessarily intertwined as the sequences we find today are the result of selection of specific instances that, when folded, minimize the energetic conflicts between their amino acids: they are overall ‘minimally frustrated’ heteropolymers \cite{bryngelson1987spin}. The energy landscape theory of protein folding recognizes these fundamental aspects and shows that the general topography of the energy landscape of globular domains is that of a rough funnel in which the native interactions are on average more favorable than non-native ones. In accordance, the population of the folding routes can be reasonably well predicted with topological models of the native state \cite{Wolynes2015} and, for most globular domains, local energetic differences rarely perturb the global aspects of the folding mechanisms \cite{ferreiro2018frustration}. This is not the typical situation in the case of repeat-proteins.

Repeat-proteins are composed of tandem arrays of similar amino acid stretches. The repeats usually fold in recursive structural elements that pack against each other in a roughly periodic way, making the overall architecture of the arrays appear as elongated objects \cite{paladin2021repeatsdb}. In these, folding domains are not easy to define and identify as several, but not necessarily all, of the repetitions co-operate in the stabilization of structures \cite{26517892}. Being quasi-one-dimensional, the folding of the complete array is dominated by the local energetics within each repeat and its local neighbors, making the folding sensitive to small perturbations that may lead to the break down of cooperativity and the appearance of stable intermediates and sub-domains \cite{26517892}. Notably, simple coarsed one-dimensional Ising-like models of repeat-protein have been found to be extremely useful for interpreting in-vitro experiments \cite{petersen2021analysis}. In general, the folding mechanisms are defined by an initial nucleation in some region of the array and the propagation of structure to their neighbors. When the local energetics are similar along the assemblage, parallel folding routes can be identified \cite{24988356}, and the routes can be switched by (de)stabilizing regions along the array \cite{notch_tripp2008rerouting}. Thus, the energy landscape of repeat-proteins appears ‘plastic’ and very amenable to design \cite{ferreiro2007plastic}. To which extent nature has exploited this opportunity is yet unknown. 

Besides single-point mutations, the evolution of repeat-proteins is thought to occur via duplications and deletions of large portions of primary structure, usually encompassing one or more repeats \cite{schuler2016evolution}. These proteins are present in all taxa and are particularly abundant in eukaryotes where they account for about 20\% of the coded proteins. Their activity is usually associated with specific protein-protein interactions, with a versatility that can be equated to that of antibodies. In various cases, the detailed folding mechanism of the repeat-arrays have been identified to play a major role in their biological function \cite{kumar2021folding}, but for most of the repeat-arrays it remains unknown. Here we aim to use evolutionary information from repeat-protein systems to investigate the folding mechanisms of thousands of natural repeat-arrays. We will make use of Ankyrin-repeat proteins as this is one of the most abundant families and their folding mechanism can be well approximated with simple folding models \cite{mello2004experimentally}. We hypothesize that the local energetics can be estimated with a maximum entropy model for the natural sequence statistics that result in a pair-wise Potts model for amino acid interactions \cite{inferring}. We map the energetics of the sequences to an Ising-model with one free-parameter that we fit with experimental folding data. The resulting model is then applied to thousands of different sequences that fold to the same overall topology, revealing distinct folding routes, the emergence of sub-domains, downhill scenarios, etc, in a variegated zoo of folding mechanisms. We found that the overall folding behavior of a complete repeat-array can be well described with few global descriptors that can be directly calculated solely from sequence information. 

\subsection{Model Definition}

\begin{figure*}
\centering
\includegraphics[width=1.\linewidth]{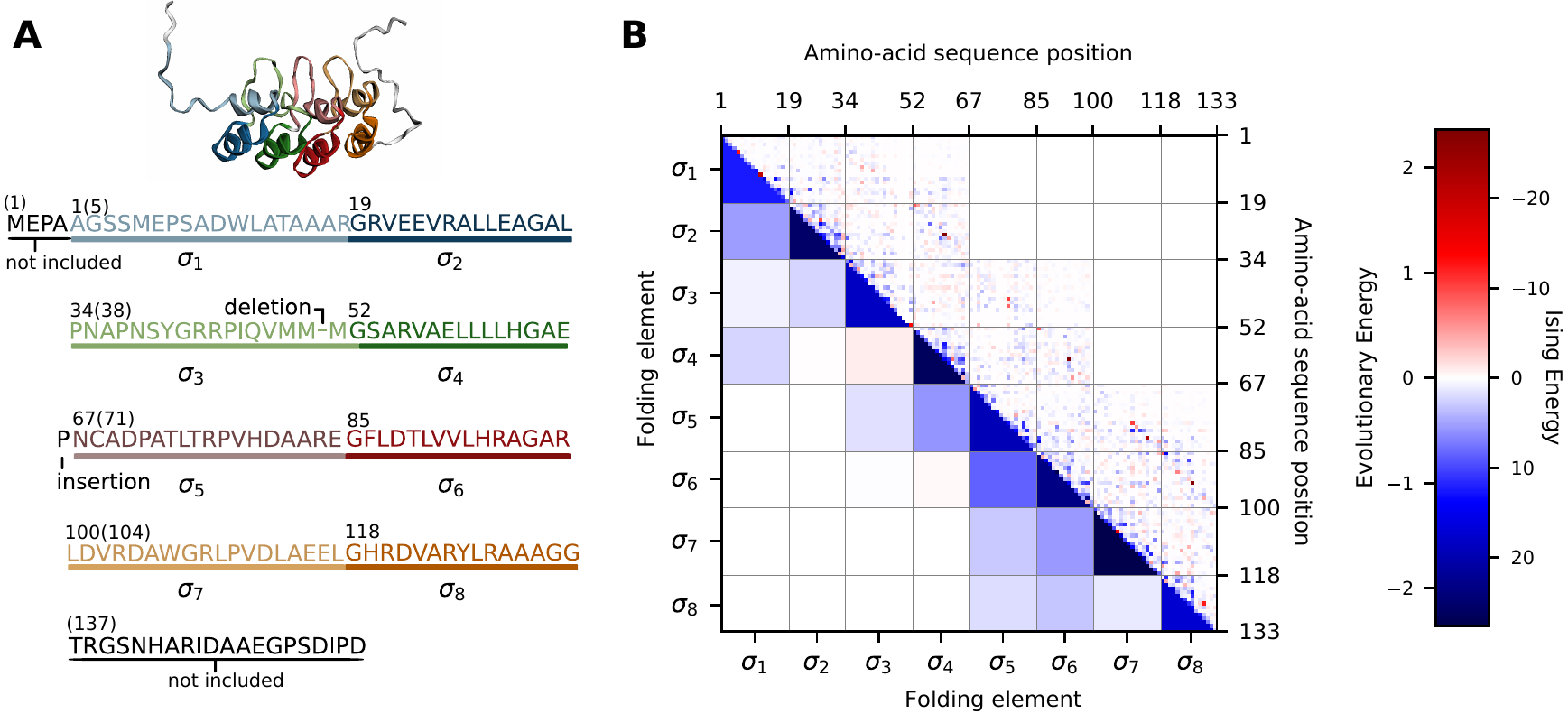}
\caption{\textbf{Model definition}. A. Folding elements are 18 and 15-residues repeat subunits as highlighted on the p16 sequence. They respectively correspond to first and second alpha-helix of each repeat as coloured on the PDB structure. B. Double energy heat-map for p16. Evolutionary residue pairwise coupĺings $\tilde{J}_{ab}(\sigma_a, \sigma_b)$ on the upper side and single residue contributions $\tilde{h}_a(\sigma_a)$ on the diagonal were added according to folding elements and re-scaled to define Ising energies $\epsilon^{i}(\bs_j)$ and $\epsilon^{s}(\bs_j,\bs_k)$ on the lower side. Blue(red) dots represent evolutionary (un)favourable pairs and positions. On the lower side, almost all internal and interaction folding energies are favourable, hence are blue.}
\label{fig:fig1}
\end{figure*}

We considered a repeat-protein as a tandem array of consecutive folding elements, each of which can be either folded (F) or unfolded (U). Each element corresponds to a group of consecutive amino-acids whose behaviour are considered together as one spin variable. The most simple assignment is to match a whole repeat with one spin, but the model can be generalised considering repeat sub-units consistently. Specific interactions take place between neighbour elements if both are folded. Therefore, the system can be represented by a finite quasi-one-dimensional folding Ising model with $N$ elements where the energy of a coarse-grained configuration, the Hamiltonian, is given by the free energy of the corresponding ensemble of microstates
 
\begin{equation}\label{eq:isingH}
    H = - \sum_{j=1}^{N}\, [T s_j \delta_{j,U}+ \epsilon^{i}_j \delta_{j,F}] - \sum_{j=1}^{N-1}\sum_{k>j} \epsilon^{s}_{jk}\delta_{j,F}\delta_{k,F},
\end{equation}

where $\delta_{j,U}$ is the Kronecker symbol, taking value 1 if element $j$ is unfolded and 0 otherwise. 
Hence, if $j$ is unfolded there is an explicit contribution to the free energy given by the entropy $s_j$ of the available spatial configurations of the element, and we take the contributions to the internal energy to be zero. If the element is folded ($\delta_{j,F}=1$ and $\delta_{j,U}=0$) we approximate the native state to be compact enough, hence there is no intrinsic entropy contribution, but an internal energy $\epsilon^i_j$ is assigned. Two elements interact with a surface energy $\epsilon^s_{jk}$ only if both are folded. 

A similar coarse-grained repeat-protein Ising model has been exhaustively studied using arbitrary parameters and compared to molecular dynamics simulations for the TPR family \cite{18483553}.  Here, we consider that the internal and surface energy parameters are functions of the amino-acid sequence. Focusing on Ankyrins, a single family in which a native repeated structure is roughly conserved \cite{Parra2015}, we hypothesise that sequence variation within repeat units is linked to changes in local stability. Hence, in order to calculate $\epsilon^i_j$ $\epsilon^s_{jk}$ for a sequence, we used co-evolutionary fields that have been inferred for the Ankyrin family using a combination of a Direct Coupling Analysis (DCA) and an explicit mechanistic evolution scheme of whole repeats duplications and deletions (details in \textit{Supporting Information}). Given a sequence, the evolutionary statistical function (often called energy) is given by a Potts model. The Hamiltonian is translational invariant and, for example, for a sequence $\bs$ with two repeats of $L$ residues it can be written as

\begin{equation}\label{eq:evo_}
    E(\bs)= - \sum_{a=1}^{2L} \tilde{h}_a(\sigma_a) -\sum_{a,b=1}^{2L} \tilde{J}_{ab}(\sigma_a, \sigma_b).
\end{equation}

This kind of inferred statistical energies have been reported to predict some fitness effect or global stability change given by point mutations \cite{contini2015many,figliuzzi2016coevolutionary,haldane2016structural,inferring}. Indeed, we compared the experimental folding energy difference between mutants and wild type ($\Delta\Delta G$) available in the literature for natural proteins of the Ankyrin family \cite{p16_tang1999stability,p16_guo2010contributions,p16_tang2003sequential,ikba_ferreiro2007stabilizing,ikba_devries2011folding,notch_street2005improved,notch_tripp2008rerouting} and we compute $\Delta E$ for the same mutants of three ANK proteins, finding a linear trend (Fig. \ref{fig:DDG}, $R^2 \simeq 0.6$). More details are provided in \textit{Supplementary Methods}. If we assume that there is no entropy difference between point mutants, the coarse-grained folding energy of fragments can be calculated simply by locally applying evolutionary fields $\tilde{h}_a$ and $\tilde{J}_{ab}$ to a sequence $\bs$ and re-scaling properly. We define $\epsilon^i_j$ and $\epsilon^{s}_{jk}$ as explicit functions of $\bs_j$ and $\bs_k$, the sequences of the folding elements $j$ and $k$  
\begin{subequations}
\label{eq:ei}
    \begin{align}
    \epsilon^{i}_j=\epsilon^{i}(\bs_j)= \frac{1}{\alpha} \Bigg[ \sum_{a\in j} \tilde{h}_a(\sigma_a) +\sum_{a,b \in j} \tilde{J}_{ab}(\sigma_a, \sigma_b) \Bigg] \\
    \epsilon^{s}_{jk}=\epsilon^{s}(\bs_j,\bs_k)=  \frac{1}{\alpha} \Bigg[  \sum_{\substack{a  \in j \\ b \in k}} \tilde{J}_{ab}(\sigma_a, \sigma_b) \Bigg] 
    \end{align}
\end{subequations}
where $a\in j$ means the sequence position $a$ is in the folding element $j$ and $\alpha=-1.3$ is the fitted slope in Fig. \ref{fig:DDG}. In this model, the internal folding energy is zero for a random sequence and is maximum for the most evolutionary favourable one.

We present p16 protein as a example of the model definitions in Fig. \ref{fig:fig1}A. We choose the folding elements to be the sequence of each alpha-helix in the typical Ankyrin repeat structure as it was done previously \cite{schafer2012discrete}. 
As in the evolution model (Eq. \ref{eq:evo_}) interactions were allowed between residues within a repeat and between first neighbour repeats, each helix in the folding model (Eq. \ref{eq:isingH}) interacts with the other one in the same repeat and with the next two. Co-evolutionary fields applied to a sequence are represented ($\tilde{J}_{ab}(\sigma_a, \sigma_b)$ as a matrix and $\tilde{h}_a(\sigma_a) $ on its diagonal) in Fig. \ref{fig:fig1}B, highlighting evolutionary favourable or unfavourable residues and pairs of residues. In this case, as in general, the partial sums of these contributions gives almost all favourable folding energy terms (Eq. \ref{eq:ei}).

The scheme we propose leaves the intrinsic configurational entropy $s_j$ undefined. We simplified its definition taking $s_j$ to be independent of amino-acid identity, therefore for an element $j$ with $L_j$ residues $s_j =L_j s$. Hence the single free parameter of the model is $s$, an average residue contribution to the effective number of configurations that a folding element polymer can access when unfolded.

\subsection{Case studies}
\begin{figure}[h]
\centering
\includegraphics[width=1.\linewidth]{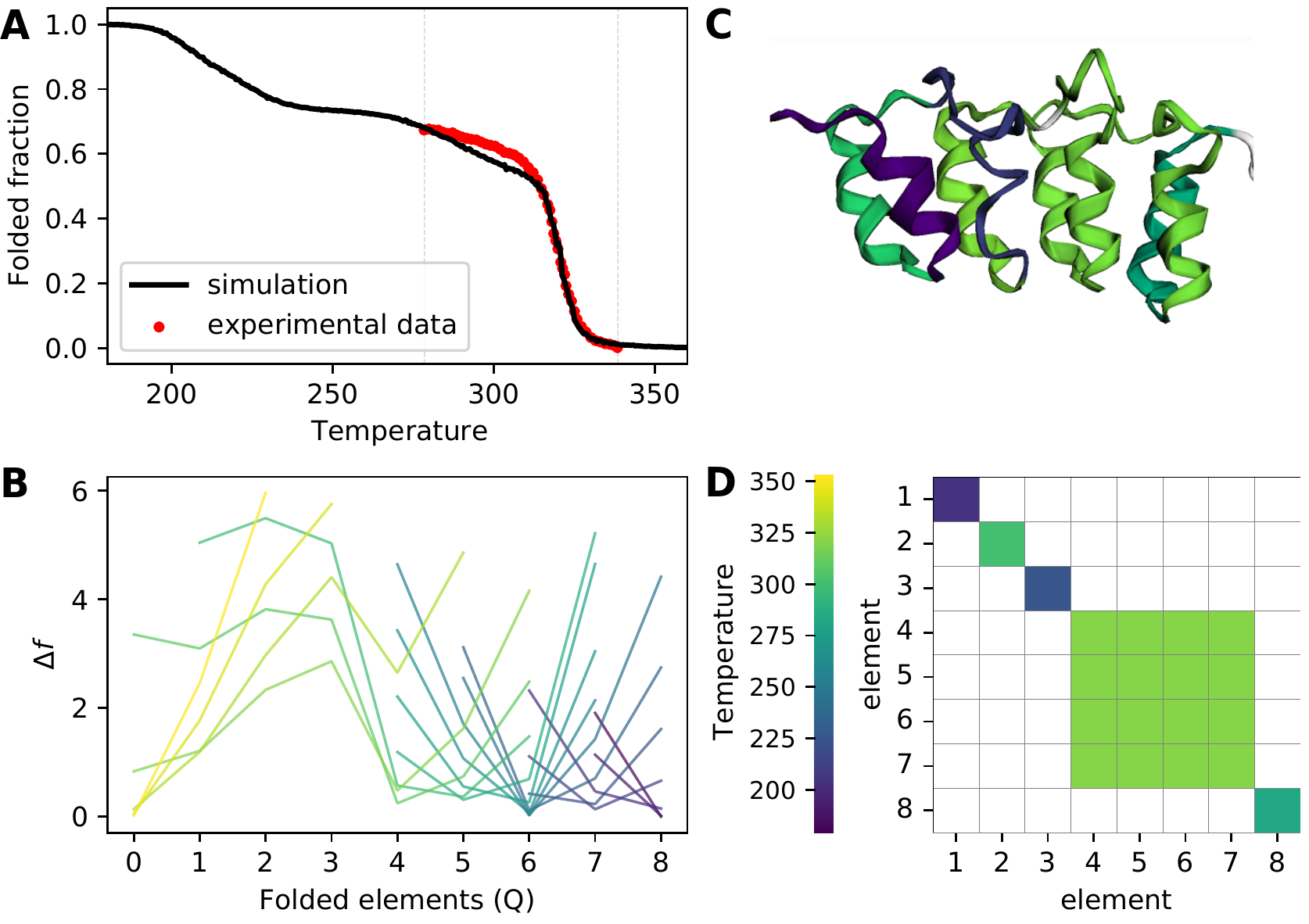}
\caption{\textbf{Simulation results for p16.} A. Simulated (black) and experimental (red) thermal unfolding curves. B. Approximate free energy profiles, coloured by temperature (same of C and D), with the number of folded elements $Q$ as reaction coordinate. There is an all-or-none transition from $Q=0$ to $Q=4$ with a barrier in between, then the minimum moves without any barrier. C. PDB structure is coloured according to the folding temperature of each element. Purple fragments are the most unstable ones. D. Apparent domain matrix, coloured by domain folding temperatures. First domain to fold (elements 4 to 7) correspond to the all-or-none transition described in B, consistently with a nucleation-propagation mechanism.}
\label{fig:fig2}
\end{figure}

We ran Monte-Carlo simulations for a well studied four-repeat ANK protein, the CDK4/CDK6 inhibitor p16. The fraction of folded elements as a function of temperature was compatible with experimental Circular Dichroism (CD) signal obtained from the literature \cite{p16_guo2010contributions} (Fig. \ref{fig:fig2}A). For the same entropy per residue $s$, the effect of a point mutation in the unfolding curve was precisely reproduced (Fig. \ref{fig:mutants}). Nevertheless, at low temperature $T$ (190-250K) simulated curves showed also a lees cooperative pre-transition. 


In the main transition, not all the elements fold together. Analysing the folding temperatures for each element, we found that elements 4 to 7 actually behave collectively in what we define as an apparent folding domain.  
Fig. \ref{fig:fig2}C-D shows that the first domain to fold (the nucleus) is closely followed by element 2 (a single-element domain). Interestingly, element 3 that behaves separately presenting low stability, correspond to the first half of the second repeat, which have been found by NMR to fold as a turn instead of a helix \cite{byeon1998tumor,yuan2000tumor}. Consistently, molecular dynamics simulations have shown in this region significant fluctuations in the folded state and it is believed to be functionally relevant for binding \cite{interlandi2006unfolding}. First and last helices also presented low folding temperatures. In addition to the border effect, we remark that the evolutionary model was learned from internal repeats, because terminal ones were considered different biological objects that can present modifications in sequence when compared to internal repeats \cite{galpern2020large,Parra2015}. 

An approximate free energy profile $\Delta f(Q)$, where $Q$ is the number of folded elements, highlights a nucleation-propagation mechanism (Fig. \ref{fig:fig2}B). The nucleation of elements 4 to 7 is an all-or-none transition from $Q=0$ to $Q=4$ with a free energy barrier in between. Structure then propagates visiting every remaining $Q$ one by one as temperature is lowered. As a measure of cooperativity, we defined a score $\rho= Q_{barrier}/(N-1) = 3/7$, the fraction of intermediary $Q$ that were not a minimum of $\Delta f(Q)$ for any $T$ in a protein with $N$ elements.

In addition to p16, we performed similar analyses on other natural ANK-containing proteins with available experimental folding data. We fit $s$ to reproduce the reversible CD thermal unfolding curves of TRPV4 \cite{trpv4_inada2012structural} , TANC1 \cite{tanc1_yang2019purification} and Kidney ANK 1 \cite{kank1_pan2018structural} finding optimal values in the range between $4.2$ and $6.2$ cal mol$^{-1}$ K$^{-1}$ res$^{-1}$ (including p16, Fig. \ref{fig:s_fit}). This interval is consistent with the calculations made by Baxa \textit{et al} ($4.3$ cal mol$^{-1}$ K$^{-1}$ res$^{-1}$ for a 18AA fragment) \cite{baxa2014loss}, it overlaps with the range estimated by D'Aquino \textit{et al} ($3.6$ to $10.5$ cal mol$^{-1}$ K$^{-1}$ res$^{-1}$) \cite{d1996magnitude} and it is lower than Makhatazde and Privalov proposal ($\sim11$ cal mol$^{-1}$ K$^{-1}$ res$^{-1}$) \cite{makhatadze1996entropy}. Given the range we obtained from experimental data fits, we set $s=5$ cal mol$^{-1}$ K$^{-1}$ res$^{-1}$ from here on, allowing us to perform simulations where thermal unfolding data is unavailable.

Consistently with reported folding dynamics \cite{ferreiro2010molecular,bradley2002limits, barrick2008folding}, our analysis on I$\kappa$B$\alpha$ (Fig. \ref{fig:ikba}) and  \textit{Drosophila melangaster} Notch Receptor (Fig. \ref{fig:notch}) allowed us to precisely identify highly cooperative domains where folding starts and distinguish them from less stable, folding-on-binding or flexible regions. On the other hand, the description of the AnkyrinR D34 24-elements fragment unfolding via a stable intermediate with an unstructured half \cite{werbeck2007probing} was not reproduced (Fig. \ref{fig:D34}). A detailed analysis of these cases is provided in \textit{Supplementary case studies}.

Although trained on natural sequences, the model can also be applied on Designed Ankyrin Repeat proteins (DARPins). Reported experimental thermal unfolding CD-signal was compatible with simulation results in two 4-repeat sequences, but  3-repeat DARPins curves were not as close to the experimental ones for the same $s$ (Fig. \ref{fig:mutants}E-F). For a library of $100$ DARPins generated using Plückthun's framework \cite{kohl2003designed} we found that stability increased with repeat-array length (Fig. \ref{fig:darpin}), consistently with previous reports \cite{wetzel2008folding}.

\subsection{General results}
\begin{figure*}
\centering
\includegraphics[width=0.8\linewidth]{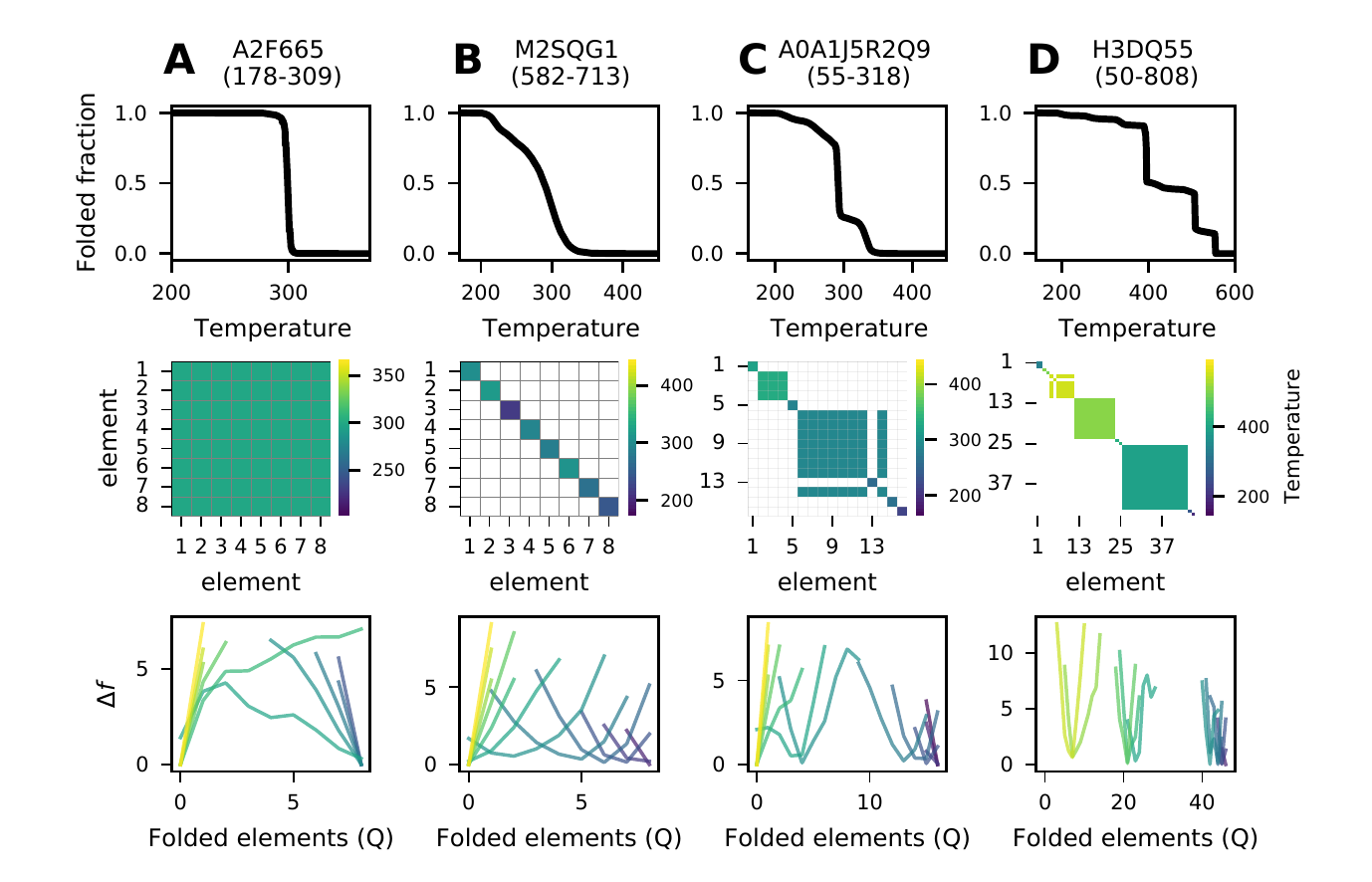}
\caption{\textbf{Different folding mechanisms.} Thermal unfolding curves (top), apparent domain matrix and temperature scale (middle) and approximate free energy profiles (bottom) for four proteins. We identified each sequence with the UniProt [ref] code and the position range of the ANK repeat-array. A. A2F665 (178-309) had a highly cooperative two-state transition and a single domain. B. M2SQG1 (582-713) did not presented any free-energy barrier, but elements unfold one by one uncooperatively. C. 8-repeat (16 folding elements) A0A1J5R2Q9 (55-318) presented many domains of different sizes and an unfolding curve with pre and post-transitions. D. H3DQ55 (50-808) is a long protein of 46 folding elements that folds in 3 steps.}
\label{fig:fig3}
\end{figure*}

To evaluate the folding of thousands of repeat-proteins, we applied the model on a \textit{selected subset} with $4020$ natural sequences formed with 4 to 36 repeats (8 to 72 elements) with no insertions or deletions. For each of these, we computed the thermal unfolding curves, the apparent domains and approximate free energy profiles and we found a large variety of folding behaviours. The data set included short proteins with strictly two-state transitions and a single domain as A2F665 (Fig. \ref{fig:fig3}A, $\rho=1$) and others like M2SQG1 (B) which presented a downhill folding without any free energy barrier ($\rho=0$). As protein size increases, we found multi-domain examples (C-D), where several nucleations and propagations appeared. Large 46-element H3DQ55 (D) presented a step-like folding, with wide stability gaps between large domain-like all-or-none transitions. Terminal repeats distinct behaviour is widespread along the data set, being more relevant in short proteins where its effect can represent up to $50\%$ of the fraction folded than in longer ones.

\begin{figure}
\centering
\includegraphics[width=0.8\linewidth]{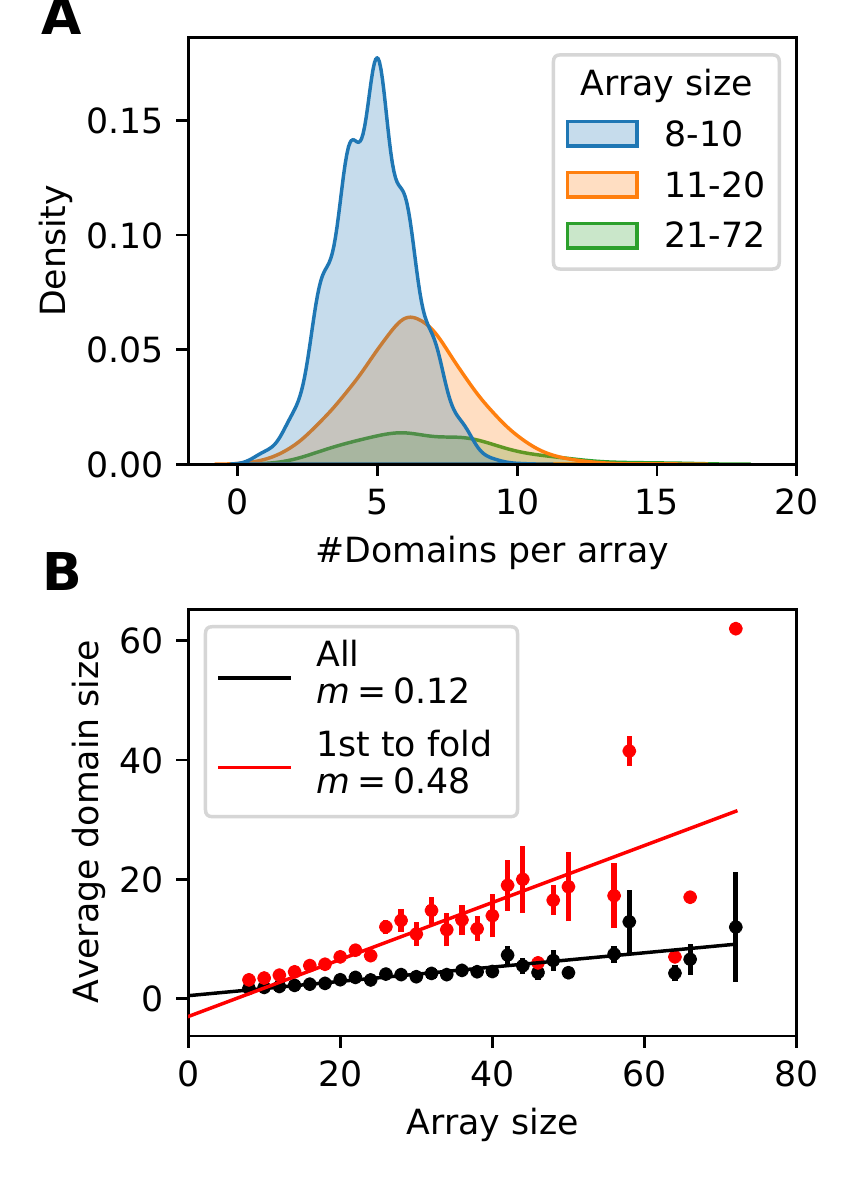}
\caption{\textbf{Domain statistics.} A. Domain count per array histogram for short (blue), intermediate (orange) and long (green) arrays. B. Average domain size as a function of array size for all (black) and only for the first domain to fold (red). Error bars are the standard errors. Linear fit slopes $m$ represented the average fraction of the array covered by a domain, which were $48\%$ for the first to fold and $12\%$ for all.}
\label{fig:fig4}
\end{figure}

There is no characteristic size for the folding domains that emerges. The proteins spontaneously fold on average 5.5 apparent folding domains, with a minor shift in the distribution when their full length is considered (Fig. \ref{fig:fig4} A). Hence, domain size grows with protein size and on average each domain covers 12\% of the tandem array: while short proteins with 8 folding elements on average present less than 2-element domains, a long array with more than 40 folding elements typically presents 10-element domains. If we consider for each protein only the first domain to fold (the nucleus), the trend with array size becomes stronger. The nucleus domain represents on average 48\% of the sequence (Fig. \ref{fig:fig4} B). Also, short domains present a wide range of folding temperatures, while longer ones (with 5 or more elements) show more stability. The nucleus domains had also a roughly exponential dependence with size, but mostly fold at physiological or higher temperatures (Fig. \ref{fig:domains_sup}).

Protein folding temperature $T_f$ and cooperativity score $\rho$ did not correlate with each other (Fig. \ref{fig:rho_tf}), suggesting the complexity of the system requires at least these two parameters to define the folding dynamics. Nevertheless, conditioning on array length reveals that there is a region of high cooperativity and stability where long arrays are more concentrated (Fig. \ref{fig:rho_tf_len}). Interestingly, it has been showed that these large repeat-proteins are naturally formed with similar and energetically favorable elements \cite{galpern2020large}. 

Ankyrin domain proteins are reported to play a variety of biological activities. We scanned all the Gene Ontology annotations with experimental evidence on the \textit{full dataset} proteins and compute for the corresponding sequences $T_f$ and $\rho$, but we did not find any clear relation between them and molecular function (Fig. \ref{fig:go}). Thus, the annotated function in the Gene Ontology is not a simple emergent of the folding properties of the sequences.

\subsection{Model interpretation}
\begin{figure*}
\centering
\includegraphics[width=0.9\linewidth]{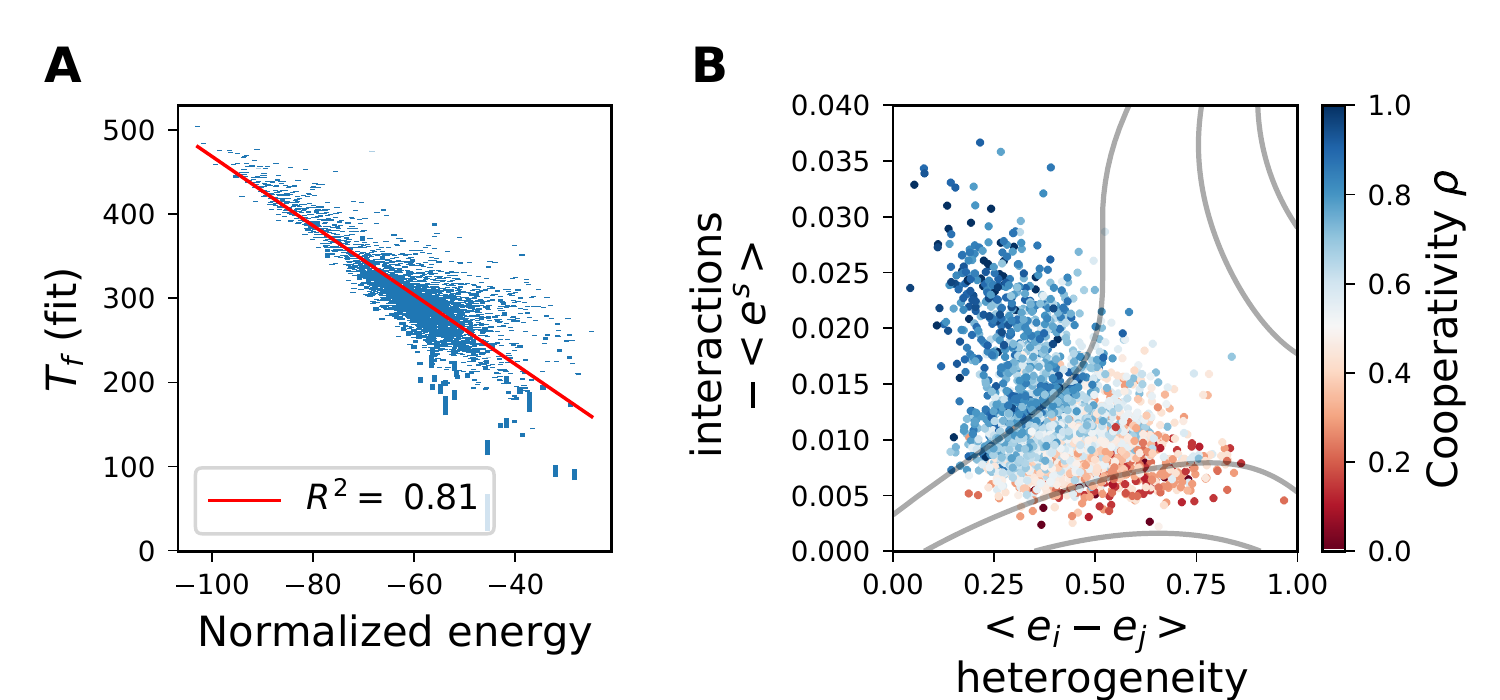}
\caption{\textbf{Model interpretation.}A. Protein $T_f$ estimations with sigmoid function fits as a function of length-normalized co-evolutionary energy of the respective sequences. B. Cooperativity score $\rho$ is showed in a color scale on a plane defined by the average normalized internal energy difference $\langle \abs{ e^i_j - e^i_k} \rangle$ and the average non-zero normalized surface energies $- \langle e^s \rangle$. Level curves (grey) were obtained with a polynomial fit.}
\label{fig:fig5}
\end{figure*}

How exactly are stability and cooperativity related to sequence statistics? On one hand, protein $T_f$ estimations were highly correlated with length-normalized co-evolutionary energy of the respective sequences Fig \ref{fig:fig5}A. This general mapping between stability and global evolutionary statistical energy was an expected output of the model, given the definitions of Eq. \ref{eq:ei} and it is consistent with experimental results for other protein families \cite{best_evo}. On the other hand, cooperativity has a more complex dependence with co-evolutionary energy. For a given protein, if differences between folding Ising internal energies $\epsilon^{i}_j$ of interacting elements (normalising by element length) were compensated by the corresponding surface energies $\epsilon^{s}_{jk}$, elements can fold cooperatively in an all-or-none transition. This would be the case of a full consensus protein with exactly duplicated repeats. But complexity arises because this is not the case for natural proteins, as one can see in Fig. \ref{fig:fig5}B. Each point corresponds to a sequence of the selected set on a plane defined by the average normalized internal energy difference $\langle \abs{ e^i_j - e^i_k }\rangle$ and the average non-zero normalized surface energies $- \langle e^s \rangle$, while the cooperativity score $\rho$ is represented by the color scale. Cooperativity smoothly changed from two-state proteins with strong surface terms and low heterogeneity to full downhill proteins on the opposite corner of the plot. In the central region, sequences could present for instance a sharp transition and a pretransition as p16 and or many barriers and intermediaries. We fitted a polynomial function of the energetic heterogeneity $\langle \abs{e^i_j - e^i_k} \rangle$ and interaction average $\langle e^s \rangle$ to define a phase diagram (Fig \ref{fig:fig5}B) and to predict cooperativity $\rho$ directly from the amino-acid sequence. 

The evolutionary model can be used to generate sequence ensembles with a Monte-Carlo run, such that they reproduced natural ANK data set features (details in \textit{Supporting Information}). For a generated ensemble of $4000$ sequences with the same protein length distribution of the \textit{selected set}, $T_f$ and $\rho$ trends with energy roughly held (Fig. \ref{fig:sup_fig5}). Hence, it is possible to generate a large ensemble of sequences and then select a subset with the desirable cooperativity and stability, without computing the Ising model.

\section{Concluding Remarks}

Although superficially simple, the energy landscapes of repeat-proteins show very rich behaviors. We explored here the folding of thousands of naturally occurring Ankyrin-repeat proteins and found that a multiplicity of mechanisms can be coded in sequences with a common low-temperature fold. On the one hand, fully cooperative all-or-none transition is obtained when the proteins are composed by sequence-similar elements and strong interactions between them. On the other hand, non-cooperative element-by-element intermittent folding becomes the rule when the elements are dissimilar and the interactions between them are energetically weak. In between these extremes, cooperative folding domains may emerge. Along the dataset 73\% of elements folds together with another ones, forming an apparent domain. Notably, we found that there is not a characteristic domain size, but we observe a scale-free domain formation. Particularly, the first domain to emerge covers about half of the repeat-array, and the rest of the chain folds upon it. We propose that the overall behaviour of the proteins can be projected in two dimensions that capture the cooperativity and the stability of the arrays. Using purely evolutionary information it is possible to predict the thermodynamics of these systems, beyond the effect of single amino acid substitutions, but the truly collective phenomena of protein folding. Strikingly, there are no examples of natural proteins for which the sequence heterogeneity is high and the interactions between them are strong. This effect can be attributed to the internal structure of the learned evolutionary field, as simulated sequences show the same distributions. Moreover, the variegated folding mechanisms is also obtained with simulated sequences, which in principle can be used to design proteins with desired folding properties and mechanical functions, for example in their nano-spring behavior \cite{lee2006nanospring}. 

The biological function of most Ankyrin-repeat arrays is thought to be mediated by specific protein-protein interactions and for many of them the folding of repeats is coupled to the binding of their targets. We identified examples for which the calculations match the known experimental region that undergoes transitions. We speculate that the rich folding behavior we identified here can be related to the biological function of these proteins, for example in the identification of the regions that undergo transitions at low temperatures as binding or allosteric regions.

Despite the recent interest on the success of structure prediction tools \cite{jumper2021highly} we should bear in mind that the dynamics of natural proteins is fundamental to their biological activity and evolution. The occupation of excited states on the energy landscapes is crucial in determining the interactions that proteins juggle in the crowded interior of cells. We presented here a way to model these dynamics solely from sequence information, that may well be applied to other types of proteins.

\section{Methods}

\subsection{Sequence data curation}
\label{sec:methods_dc}
We used subsets of a 1.2 million Ankyrin repeat sequences alignment, previously built and characterised \cite{galpern2020large}. More details are provided in \textit{Supplementary methods}.

\subsection{Evolutionary model for repeat arrays}
We used the evolutionary energy fields learned with a statistical model from an ANK multiple sequence alignment (MSA), which included internal but not terminal repeats from the same full dataset previously described \ref{sec:methods_dc} and arrays up to 40 repeats long. Briefly, the model combines Direct Coupling Analysis (DCA) and an explicit evolution mechanism of duplications and deletions of repeats. Maximum entropy model uses as constraints empirical MSA single site and 2-point amino-acid frequencies, pairwise repeat identity and array length distribution. Statistical inference was made with a Boltzmann Learning algorithm, obtaining energy fields $\tilde{h}_a(\sigma_a)$ and $\tilde{J}_{ab}(\sigma_a, \sigma_b)$. More details about the model definition, parameter inference and validation are provided in \textit{Supporting Information}).

\subsection{Ising model elements assignment}
The multiple repeat sequence alignment we worked with has the amino-acid pattern \texttt{TPLH} on positions 10 to 13. We divided each repeat in two fragments using secondary and tertiary structural information reported in the literature \cite{sedgwick1999ankyrin}. We labelled residues 1-18 containing $\beta$-hairpins and a $\alpha$-helix as the fragment $A$ and residues 19-33 containing the second $\alpha$-helix as the fragment $B$. Given that the dataset only contained concatenated 33 amino-acid repeats, any repeat-array can be mapped to a periodic succession of $A$-$B$ elements. Repeats with less than 33 residues include gaps ('-') to complete all the 33 positions. Although the evolutionary field was learned including gaps in the alphabet and rigorously there is a folding energy contribution to be considered, we set both energetic and entropic gap positions contribution to zero.

\subsection{Ising model implementation}
We performed Monte Carlo Metropolis algorithm simulations of the finite Ising model with a python routine.
Simulation total time, transient time and equilibration time parameters scale linearly with protein length, and were obtained with an autocorrelation analysis. Deletions, unknown amino-acids and missing residues at the beginning or ending of sequences were excluded from all calculations.
For the \textit{selected dataset}, simulations were made for 500 equispaced temperatures in an interval such that the system folds and unfolds completely.


\subsection{Free energy profiles approximation}
We obtained qualitative free energy profiles approximating the probability of states $s$ with $Q$ folded elements with the Metropolis Monte-Carlo sampling. We considered together sampled states for simulations performed in a window of the 10 closest temperatures. The profiles we used are computed as 

\begin{equation}
    \Delta f (Q) =- kT \, log \left( \frac{\sum_{\substack{s|Q}}N(s)}{\sum_s N(s)}\right),
\end{equation}
where $T$ is the average temperature, $N(s)$ are the counts of state $s$ and $s|Q$ are the states with $Q$ folded elements. 


\subsection{Apparent domains}
Elements $j$, $k$ were assigned to the same domain if $\abs{T_f^j-T_f^k}<5$, where the folding temperature $T_f^j$ was obtained by a sigmoid fit of the folding probability of element $j$. Domain folding temperature is the average $\langle T_f^j \rangle$ for $j$ belonging the domain. Overlapping domains were separated into the minimum number of non-overlapping ones. If more than one separation is possible, temperature difference between domains were maximised.

\subsection{Folding temperature}
To fit folding temperatures $T_f$ we approximated the fraction folded $m (T)$ as  
\begin{equation}
    m (T) =   \frac{ m_{max} }{ 1 + e^{a(T-T_f)}}
\end{equation}
where $m_{max} \in [0,1]$. We used \texttt{scipy} library \texttt{curve\_fit} to fit and get $\sigma_{T_f}$, which we used as $T_f$ errors. 

\subsection{Cooperativity phase diagram fit}
We made multivariate polynomial fits on the \textit{selected set}, making 5-fold cross-validation and using python \texttt{sklearn} library \texttt{preprocessing.PolynomialFeatures} and \texttt{linear\_model.LinearRegression}. We compared performance measured by predicted $\rho$ RMSE for 1 to 9-degree polynomial and kept the best one, 3-degree.


\medskip

{\bf Acknowledgements. 
} This work used computational resources from CCAD – Universidad Nacional de Córdoba (\url{ https://ccad.unc.edu.ar/}), which are part of SNCAD – MinCyT, República Argentina. This work was supported by the Consejo de Investigaciones Científicas y Técnicas (CONICET); the Agencia Nacional de Promoción Científica y Tecnológica [PICT2016-1467 to D.U.F.] and Universidad de Buenos Aires (UBACYT 2018 - 20020170100540BA). Additional support from NAI and Grant Number: 80NSSC18M0093 Proposal: ENIGMA: EVOLUTION OF NANOMACHINES IN GEOSPHERES AND MICROBIAL ANCESTORS (NASA ASTROBIOLOGY INSTITUTE CYCLE 8).
We thank I. E. S\'anchez and P. G. Wolynes for stimulating discussions and comments during the development of this work. 
\medskip

\newpage
\beginsupplement
\section{Supporting Information Text} 

\subsection{Evolutionary model for repeat arrays}
\subsubsection{Evolutionary model}\label{sec:repevo_model}

In the evolutionary model for repeat-proteins, introduced in~\cite{thesis_jacopo}, we consider an array of $N_r$ repeats in tandem, each consisting  of an amino-acid sequence of fixed length $l_r$. 
Repeats are duplicated and deleted with deletion and duplication rates  per repeat that we assume to be equal $\mu_{dup}=\mu_{del}$, captured by a unique parameter $\mu_d$.   The rate at which these event happen at the whole array level depends linearly on the array length, so that the overall array duplication (and deletion) rate is $\mu_d N_r$. Here duplications always place repeats one next to each other conserving the repeat locality on the array.  This dynamics determines the phylogenetic relationship between repeats and map it to their relative position in the array, alongside mutations which spark mismatches along this repeat phylogeny. 

These size changes undergo selection $S(N_r)$, defined as the probability that a size change leading to an array of length $N_r$ is accepted.
We assume that $S$ depends only on the number of repeats in the array and not on the amino-acid sequence.
The master equation for the probability of $N_r$ is 
\begin{equation}\label{PN}
\begin{split}
    \frac{dP(N_r)}{dt} = & (P(N_r - 1) (N_r-1) + P(N_r + 1) (N_r+1)) S(N_r) \mu_{dup} - \\
    & P(N_r) N_r (S(N_r - 1) + S(N_r + 1)) \mu_{del}
 \end{split}
\end{equation}
where we set $S(N_r)$ so that the equilibrium distribution matches the empirical array length distribution $P^{emp}(N_r)$ in our dataset.

\begin{figure}
\includegraphics[width=1.0\linewidth]{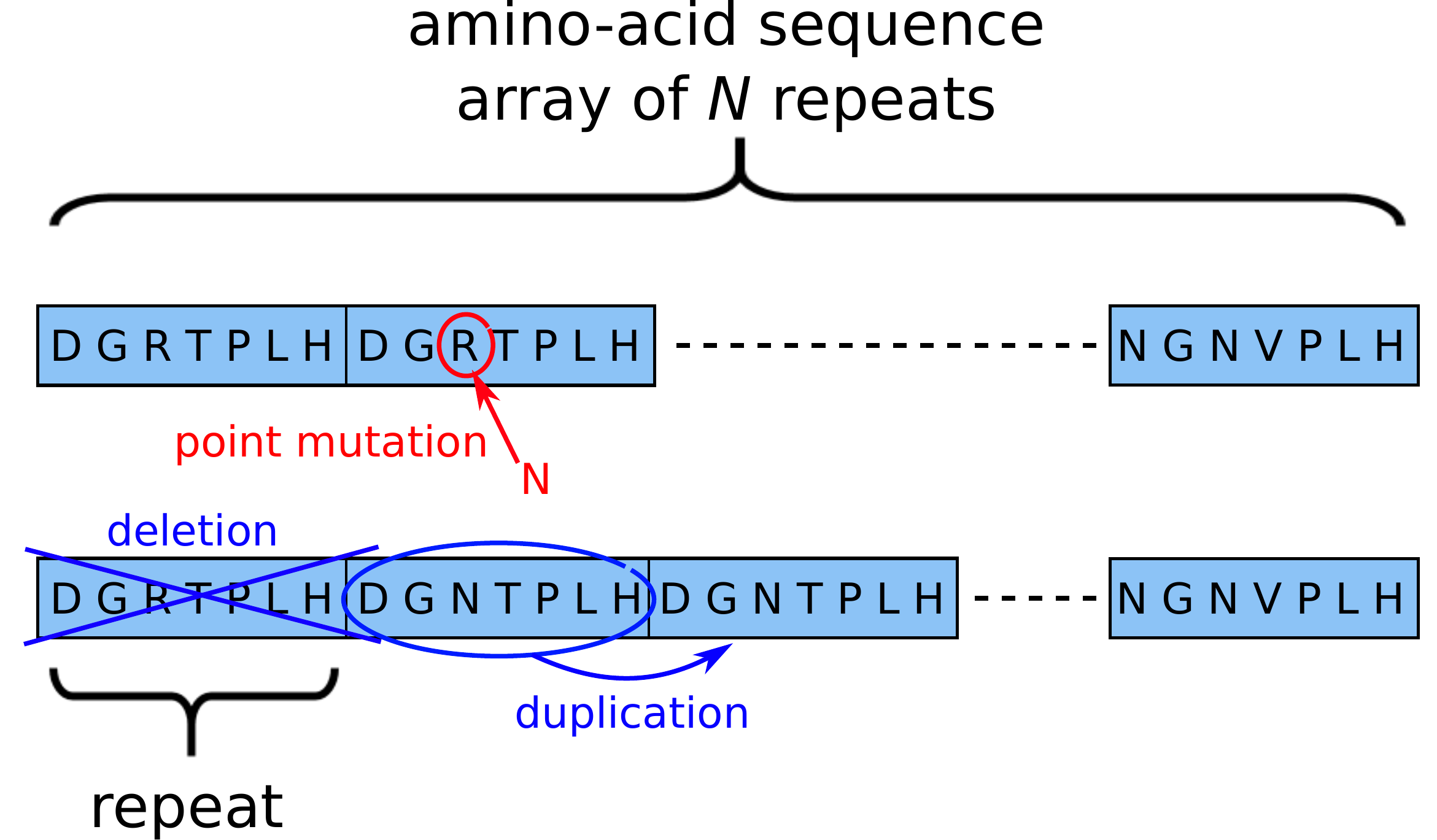}
\caption{ Key events characterizing our model for repeat tandem arrays evolution. Within an array many repeat duplications and deletions of whole repeats happen at rate $\mu_d$, whereas point mutation happen at rate  $\mu_p$.
\label{cartoon_dupdelmut:fig}
}
\end{figure}

Point mutations can occur with rate  $\mu_p$ per amino-acid site, which with duplications and deletions constitute the key events underlying this simple model (fig.~\ref{cartoon_dupdelmut:fig}).
After mutations, sequences undergo selection according to some evolutionary energy. 
This energy is defined by an internal repeat Potts energy 
\begin{equation}
E_1(\bs) = - \sum_{i=1}^{l_r} h_i(\sigma_i) -\sum_{i<j\leq l_r} J_{ij}(\sigma_i, \sigma_j) 
\end{equation}
 acting on single repeats separately, plus  an interaction term $I^{i, i+1}$ between consecutive repeats $i, i+1$, consisting of  coupling $J$s between repeats. 
For example for a 2 repeats array we have:
\begin{equation}\label{H_2} 
\begin{split}
	 E_2(\bs)= & \underbrace{- \sum_{i=1}^{l_r} h_i(\sigma_i) -\sum_{i<j\leq l_r} J_{ij}(\sigma_i, \sigma_j)}_{E_1(\bs^1) = E_1^1} \\
	  &\underbrace{ - \sum_{i= 1}^{l_r} h_i(a_{i + l_r}) - \sum_{i<j \leq l_r} J_{ij}(a_{i + l_r}, a_{j + l_r})}_{ E_1^2} \\ 
	  &\underbrace{-\sum_{i<l_r \leq j} J_{ij}(\sigma_i, \sigma_j)} _{I^{1,2}} 
\end{split}
\end{equation}
With these objects we can generalize a discrete translational invariant energy for arrays of arbitrary length $N_r$:
\begin{equation}\label{H_N} 
 E_{N_r}(\bs)= \sum_{i=1}^{N_r}E_1^i + \sum_{i=1}^{N_r - 1} I^{i, i+1}\,.
\end{equation}

This minimal model assumes independent selection on protein lengths and sequences --- modulo boundary effects due to the fact that the impact of $I^{1,2}$ is lower on terminals. Apart from the energy parameters, the only relevant parameter is the ratio of the two rates, $\mu_r = \frac{\mu_{d}}{\mu_{p}}$, or equivalently the ratio between the two timescales $t_r = \frac{1}{\mu_r} = \frac{\< t_{d} \>}{\< t_{p}\>}$. 

As a side note, we mention that  this model is out of equilibrium because microscopically detailed balance is broken by duplications and deletions. Repeat arrays are characterized by joint probability $P(\bs, N_r)$ for the sequence $\bs$  with $N_r$ repeats. 
If  $\frac{\langle t_{d} \rangle}{\langle t_{p} \rangle} \gg  1 $ we have a separation of timescales between the two processes, so that the mutation process can thermalize between typical dupdel times and multi-repeat amino-acid sequences are almost always at equilibrium with $P(\bs | N_r) \rightarrow (1/Z_{N_r})e^{-E_{N_r}(\bs)}$. 

\subsubsection{Numerical simulations}\label{sec:repevo_SI_sim}

In order to simulate our model we use a Metropolis-Hastings Monte Carlo scheme both for changes in array length and in sequence. Therefore  we start from a random amino-acid sequence and we  produce point mutations with rate $\mu_p=1$ per site, one at a time. If a mutation decreases the evolutionary energy \eqref{H_N} we accept it. Otherwise we accept the mutation with probability $e^{-\Delta E}$, where $\Delta E$ is the difference  of energy between the original and the mutated sequence.  

In parallel, we generate duplications and deletion events with rate $\mu_d$ per repeat, therefore with absolute rate $\mu_d  N_r$, producing a change of array length $N_r^o \rightarrow N_r^n = N_r^o \pm 1$. The resulting array is accepted with probability 
\begin{equation}
    acc(N_r^o \rightarrow N_r^n)= \min \left(1,\frac{S(N_r^n) N_r^n}{S(N_r^o) N_r^o}\right),
\end{equation}

We estimate the order of magnitude of the limiting process of the model as $t_{lim} = \max{(10 \<t_p\>, \<t_d\>)}$. We skip the first $100 t_{lim}$ sequences, and then we add a sequence every $t_{lim}$ to the final ensemble after checking that these conditions are sufficient to reach thermalization and to draw sequences that are uncorrelated in both processes.

\subsubsection{Parameters inference}\label{ssect:repevo_inference}

In order to obtain a model that reproduces the experimentally observed site-dependent amino-acid  frequencies, $f_i(\sigma_i)$ and correlations between two positions $f_{ij}(\sigma_i, \sigma_j)$ within a single repeat and between consecutive repeats,  we apply a likelihood gradient ascent procedure, starting from an initial guess of the $ h_i(\sigma_i)$, and $J_{ij}(\sigma_i, \sigma_j)$ parameters. At the same time we learn $\frac{\langle t_{d} \rangle}{\langle t_{p} \rangle} $  to reproduce the average similarity (number of matching amino-acids) between neighboring repeats, $\< ID_{1st}^{\rm emp} \> $. This extra optimization step leaves the learning problem convexity unaffected, because the model $\< ID_{1st} \>$ depends monotonously on the scalar parameter $\mu_r$: the higher $\mu_r$ the higher  $\< ID_{1st} \>$. 
This monotonic trend implies that  the proper update direction for $\mu_r$  is proportional to $ \< ID_{1st}^{\rm emp} \> - \< ID_{1st}^{\rm model} \>$ .
Importantly, learning a full field for pairs of consecutive repeats without the dynamic ingredient of dupdels cannot reproduce this distribution or its average (not shown) as was already found in~\cite{Espada2018} for a different dataset.

At each step, we  generate 150000 sequences of variable length through the Metropolis-Hastings Monte-Carlo sampling
described in~\ref{sec:repevo_SI_sim}. 
Once we have generated the sequence ensemble, we measure its marginals $f^{\rm model}_i(\sigma_i)$ and $f^{\rm model}_{ij}(\sigma_i, \sigma_j)$, the latter at most between consecutive repeat pairs, as well as $\< ID_{1st}^{\rm model} \> $,
and update the parameters of  Eq.~\ref{H_N} and  $\frac{\langle t_{d} \rangle}{\langle t_{p} \rangle} $ following the gradient of the likelihood, equal to the difference between model and data averages.
In order to speed up the inference we
add an inertia term to the gradient ascent mimicking acceleration, as described in~\cite{goh2017why}.

The local field is updated as:
\begin{equation}\label{h_upd_repevo_SI} 
h_i(\sigma_i)^{t+1} \leftarrow h_i(\sigma_i)^{t} + \epsilon_m[f_i(\sigma_i) - f_i^{\rm model}(\sigma_i)] \\
+ I_{tot}  (h_i(\sigma_i)^{t}  - h_i(\sigma_i)^{t-1} ),
\end{equation}
As the number of parameters for the interaction terms $J_{ij}$ is large, we impose a sparsity constraint via a $L_1$ regularization $\gamma \sum_{ij,\sigma,\tau}|J_{ij}(\sigma,\tau)|$ added to the likelihood. This leads to the following rules of maximization:\\
If $J_{ij}(\sigma_i, \sigma_j)^{t} = 0 $ and $\abs{f_{ij}(\sigma_i, \sigma_j) - f_{ij}^{\rm model}(\sigma_i, \sigma_j)} < \gamma$
\begin{equation}\label{j_upd1} 
J_{ij}(\sigma_i, \sigma_j)^{t+1} \leftarrow 0.
\end{equation}
If $J_{ij}(\sigma_i, \sigma_j)^{t} = 0 $ and $\abs{f_{ij}(\sigma_i, \sigma_j) - f_{ij}^{\rm model}(\sigma_i, \sigma_j)} > \gamma$
\begin{equation}\label{j_upd2} 
\begin{split}
J_{ij}(\sigma_i, \sigma_j)^{t+1} \leftarrow   \epsilon_j[ & f_{ij}(\sigma_i, \sigma_j) - f_{ij}^{\rm model}(\sigma_i, \sigma_j) - \\
& \gamma {\rm sign}(f_{ij}(\sigma_i, \sigma_j) - f_{ij}^{\rm model}(\sigma_i, \sigma_j))].
\end{split}
\end{equation}
If $\Big[J_{ij}(\sigma_i, \sigma_j)^{t} + \epsilon_j[f_{ij}(\sigma_i, \sigma_j) - f_{ij}^{\rm model}(\sigma_i, \sigma_j) - \gamma {\rm sign}(J_{ij}(\sigma_i, \sigma_j)^{t})]\Big]  J_{ij}(\sigma_i, \sigma_j)^{t} \geq 0 $ 
\begin{equation}\label{j_upd3} 
\begin{split}
J_{ij}(\sigma_i, \sigma_j)^{t+1} \leftarrow  & J_{ij}(\sigma_i, \sigma_j)^{t} + \epsilon_j[ f_{ij}(\sigma_i, \sigma_j)  \\
& - f_{ij}^{\rm model}(\sigma_i, \sigma_j) -  \gamma {\rm sign}(J_{ij}(\sigma_i, \sigma_j)^{t})] \\
  & + I_{tot}(J_{ij}(\sigma_i, \sigma_j)^{t}  - J_{ij}(\sigma_i, \sigma_j)^{t-1} ).
\end{split}
\end{equation}
If $\Big[J_{ij}(\sigma_i, \sigma_j)^{t} + \epsilon_j[f_{ij}(\sigma_i, \sigma_j) - f_{ij}^{\rm model}(\sigma_i, \sigma_j) - \gamma {\rm sign}(J_{ij}(\sigma_i, \sigma_j)^{t})]\Big]  J_{ij}(\sigma_i, \sigma_j)^{t} < 0 $ 
\begin{equation}\label{j_upd4} 
J_{ij}(\sigma_i, \sigma_j)^{t+1} \leftarrow  0.
\end{equation}
The times ratio parameter $t_r=\frac{\langle t_{d} \rangle}{\langle t_{p} \rangle} $ is updated according to:
\begin{equation}\label{t_upd} 
t_r^{t+1} \leftarrow t_r^{t}  - \epsilon_{\rm ID}(\< ID_{1st}^{\rm emp} \> - \< ID_{1st}^{\rm model} \>).
\end{equation}

To estimate the model error, we compute  $f_i(\sigma_i) - f_i^{\rm model}(\sigma_i)$, $f_{ij}(\sigma_i, \sigma_j) - f_{ij}^{\rm model}(\sigma_i, \sigma_j)$. 
We repeat the procedure above until the maximum of all errors, $|f_i(\sigma_i) - f_i^{\rm model}(\sigma_i)|$, $|f_{ij}(\sigma_i, \sigma_j) - f_{ij}^{\rm model}(\sigma_i, \sigma_j)|$ , goes below $0.004$. The order of magnitude of this threshold value is motivated by the finite size effects from the number of samples in our dataset. The empirical frequencies can be thought as the frequencies of the result of $N_s$ Bernoulli trials (each of this trials draws the symbols in a sequence of our sample), and therefore they are distributed according to a multinomial distribution parametrized by some underlying true distribution $p(\sigma)$. The standard deviation of these measured frequencies will be of order $O(\frac{1}{\sqrt{N_s}})$ that for example for the $\sim 420000$  repeats in our dataset is $\sim 0.002$, therefore on the same order of magnitude of our threshold.  
Moreover we require that $|\< ID_{1st}^{\rm emp} \> - \< ID_{1st}^{\rm model} \>| < 0.1$, and this other threshold  scale is derived from the empirical difference between first and second neighbors similarity  $|\< ID_{1st}^{\rm emp} \> - \< ID_{2nd}^{\rm emp} \>| = 0.4$.

Using this procedure we  calculate the model defined in Eq.~\ref{H_N} with different interaction ranges $W$ for the couplings $J_{ij}$, exactly as we did in~\cite{marchi2019_ploscb}.
 We  start from the independent model $h_i(\sigma_i)= \log{f_i(\sigma_i)}$. We first learn the model in  Eq.~\ref{H_N}  with $J=0$, that consists of just learning $t_r$. 
We then re-learn models with interactions between sites $i,j$ along the linear sequence such that $\abs{i -j}\leq W$, 
in a seeded way starting from the previous model.
We progressively increase $W$ until we reach the full repeat pairs model, $W=66$.

The optimization parameters were set to  $\epsilon_{\rm ID}=0.5$, and $\gamma=0.0003$, while $\epsilon_m \in [0.1,1]$, $\epsilon_j  \in [0.05,1]$, and $ I_{tot}  \in [0.7,0.95]$ were tuned ad-hoc as a function of $W$, the first two in a decreasing fashion.

\subsubsection{Inference validation}\label{sec:repevo_res_basic}

\begin{figure}
\centering
\includegraphics[width=1.0\linewidth]{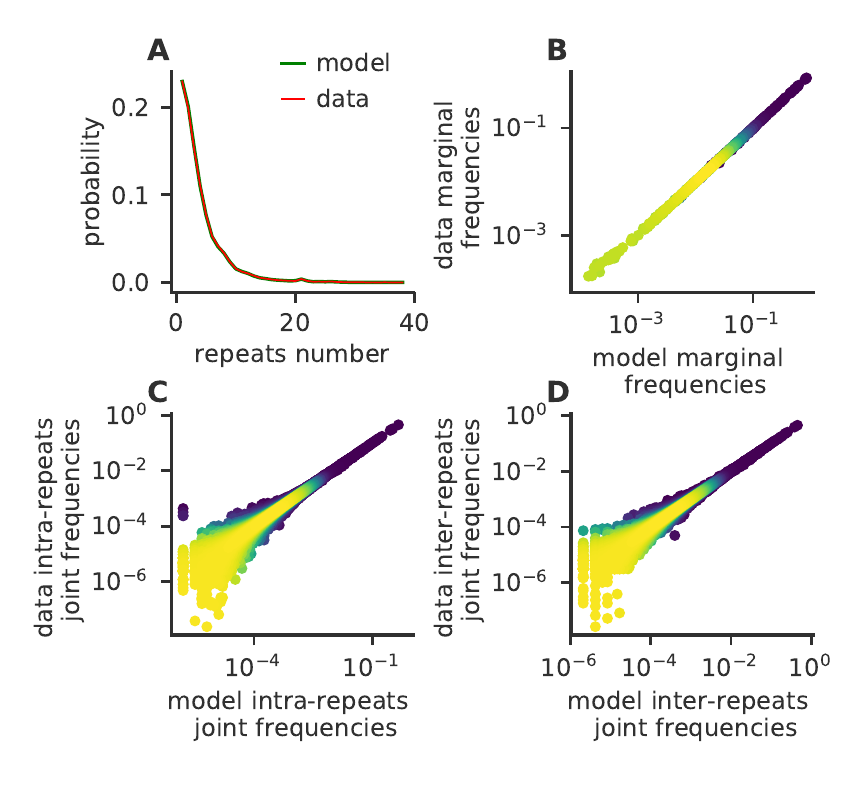}
\caption{ The inferred model reproduces the desired empirical statistics.
A) Probability distribution of number of repeats in an array, data in red and model generated sequences in green. 
B), C), D) Scatter plot between the empirical and model generated 1 site marginal amino-acid  frequencies,  2 site joint amino-acid frequencies within the same repeat, 2 site joint amino-acid frequencies between consecutive repeats, respectively. The color map represents points density (yellow higher density).
\label{check_stat_learn:fig}
}
\end{figure}

As a first consistency check we show in fig.~\ref{check_stat_learn:fig} that the inferred model reproduces both the empirical array length distribution (panel A) and the three amino-acid frequencies sets (panels B,C,D respectively) we used to fit the parameters.  Thanks to our inference scheme we also learn quantitatively the timescales ratio between dupdels and mutations: $\frac{\< t_d\>}{\<t_p\>} = \frac{1}{\mu_r} = 27.28 $. Therefore on average duplications (the average time for deletions is the same) per repeat happen $27.28$ times slower than mutations per site. Putting times in the same relative scale, per repeat, replacing $\< \tilde t_p\>  = \frac{ \< t_p\>}{l_r} = \frac{ \< t_p \>}{33} $  we have   $\frac{\< t_d\>}{\< \tilde t_p\>}  \sim 900 $, therefore duplications are about 3 orders of magnitude rarer than mutations and the system is almost at equilibrium. This is consistent with the fact that this procedure reproduces the right equilibrium distribution shown in  fig.~\ref{check_stat_learn:fig}.

\begin{figure}
\centering
\includegraphics[width=1.0\linewidth]{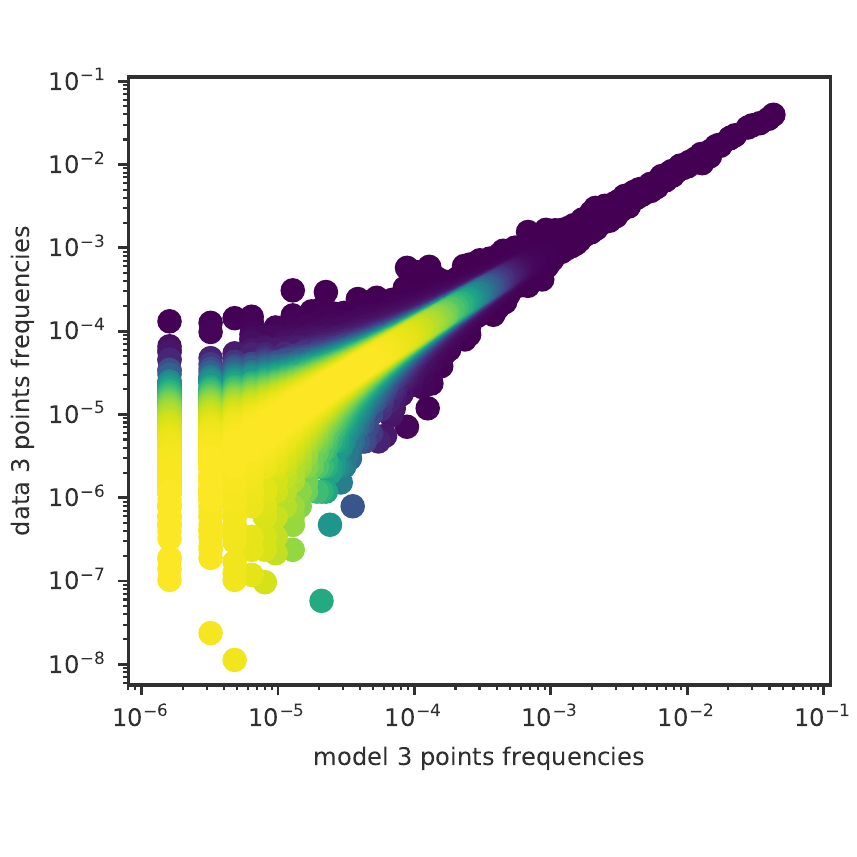}
\vspace*{-10mm}
\caption{ Scatter plot between the empirical and model generated 3 sites joint amino-acid frequencies, the color map represents points density (yellow higher density). The model reproduces well higher order statistics that were not used for fitting.
\label{3points:fig}
}
\end{figure}

Fig.~\ref{3points:fig} shows that also 3 point amino-acid frequencies $f_{i,j,k}(\sigma_i, \sigma_j, \sigma_k)$, including inter-repeat sites, are well reproduced by the model, even though we did not use them to infer the model. This means that the model generalizes well with respect to some higher order statistics that were not used for fitting.

\subsubsection{Simplified notation used in the main text}
Given $N_r$ it is possible to consistently repeat in a single matrix all the local contributions $h_i$ by defining  
\begin{equation}
    \tilde{h}_a=\tilde{h}_{i+n\,l_r}=h_i,
\end{equation}
and for $J_{ij}$ contributions by defining
\begin{equation}
    \tilde{J}_{ab}=\tilde{J}_{i+n\,l_r,j+n\,l_r}=J_{ij},
\end{equation}
where $n$ goes from $0$ to $N_r-1$.


\subsection{Supplementary case studies}

\subsubsection{I$\boldsymbol{\kappa}$B$\boldsymbol{\alpha}$} 
\begin{figure}
\centering
\includegraphics[width=1.\linewidth]{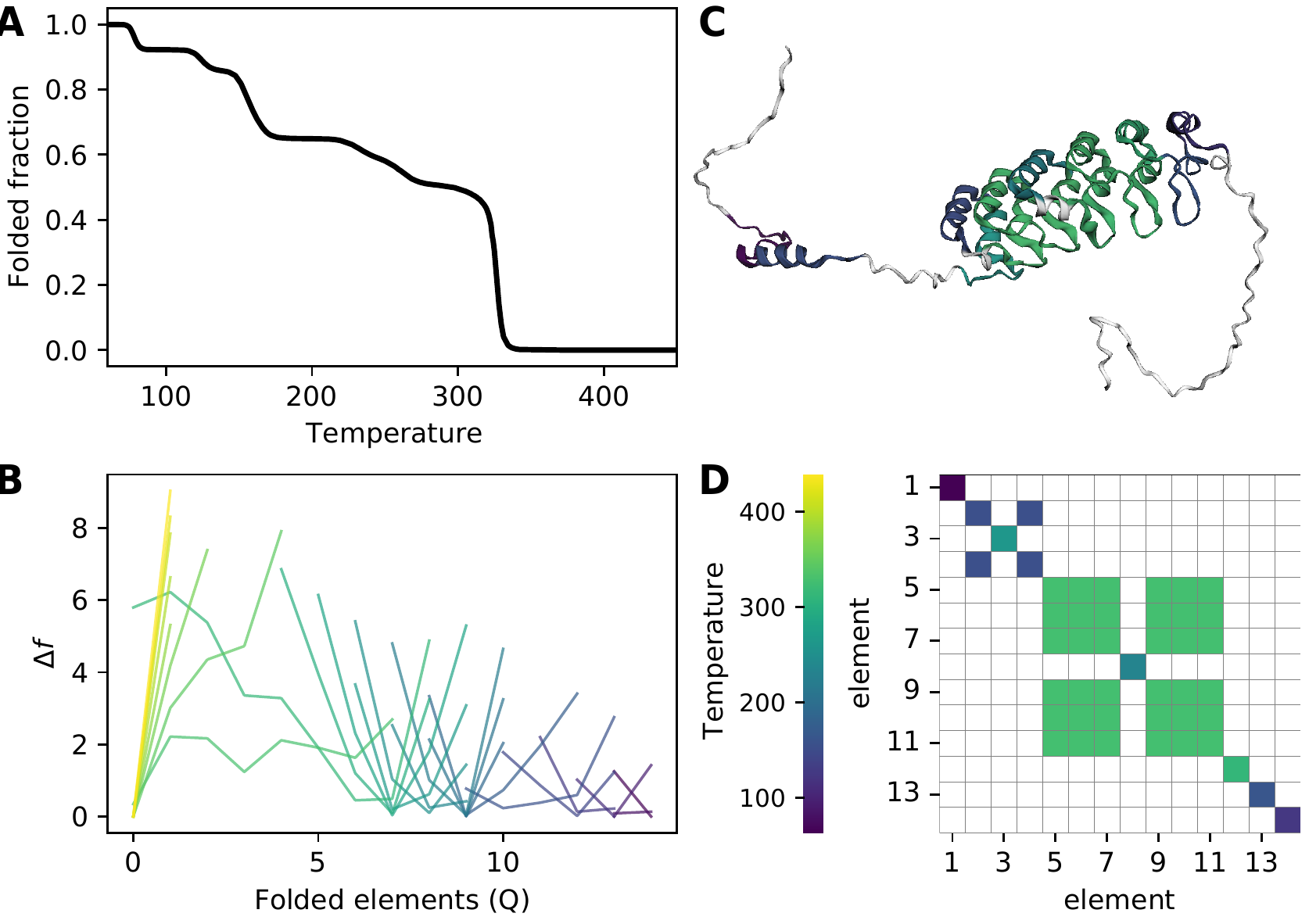}
\caption{\textbf{Simulation results for I$\boldsymbol{\kappa}$B$\boldsymbol{\alpha}$.} A. Thermal unfolding curve. B. Approximate free energy profiles, coloured by temperature (same of C and D), with the number of folded elements $Q$ as reaction coordinate. C. ALphaFold \cite{alphafold} structure is coloured according to the folding temperature of each element. Purple fragments are the most unstable ones. D. Apparent domain matrix, coloured by domain folding temperatures.}
\label{fig:ikba}
\end{figure}

We consider a 7 ANK repeat fragment (including residues 23-55  not considered in previous studies as a first ANK repeat), that is to say a tandem array of 14 elements.
Simulations show a cooperative transition of the most stable elements: 5, 6 and 8 to 12, which are followed by less cooperative folding of the remaining ones (Fig. \ref{fig:ikba}). This description is compatible with denaturation experiments, where a major cooperative event were followed by a non-cooperative one mapped to elements 11 to 14, a region that was also reported to fold upon binding \cite{ferreiro2010molecular}. Point mutant Y254L-T257A do not fully stabilised the C-terminal repeat (elements 13 and 14) as been characterised \cite{ferreiro2010molecular} but locally increase the folding temperature of element 13, C-terminal first helix, in our simulations (data not shown). We detected element 7 as less stable than others, while NMR studies have shown that the region of elements 7 and 8 are high flexible in the folded state and are involved in binding \cite{ferreiro2010molecular}. Finally, element 1 is the last fragment to fold in simulations and consistently has no alpha-helix structure according to an Alpha-Fold prediction.

\subsubsection{Notch receptor} 
\begin{figure}
\centering
\includegraphics[width=1.\linewidth]{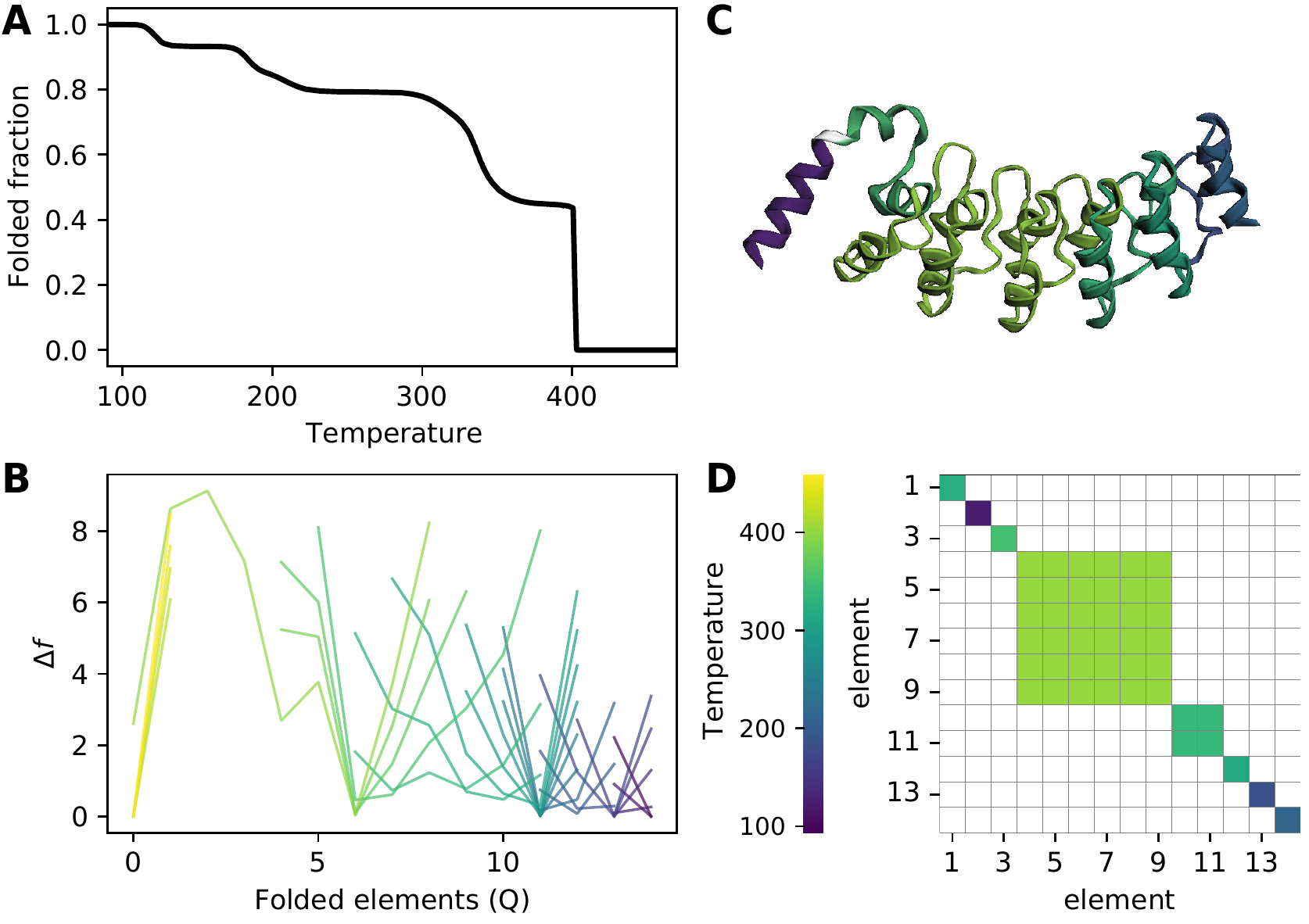}
\caption{\textbf{Simulation results for \textit{Drosophila melangaster} Notch receptor Ankyrin Domain.} A. Thermal unfolding curve. B. Approximate free energy profiles, coloured by temperature (same of C and D), with the number of folded elements $Q$ as reaction coordinate. C. PDB structure is coloured according to the folding temperature of each element. Purple fragments are the most unstable ones. D. Apparent domain matrix, coloured by domain folding temperatures.}
\label{fig:notch}
\end{figure}

We analysed the \textit{Drosophila melangaster} Notch receptor Ankyrin region, a 14-elements array. We find that the first 3 elements are less stable and do not fold cooperatively with the next ones (Fig. \ref{fig:notch}). Consistently, experimental evidence suggest that the region of the first two elements remains at least partially folded while the main subdomain is fully folded \cite{bradley2002limits}. We detect that the nucleus domain are elements 4 to 9, while the next elements showed less stability, another observation in agreement with dynamic data and previously proposed models \cite{bradley2002limits}.

\subsubsection{D34} 
\begin{figure}
\centering
\includegraphics[width=1.\linewidth]{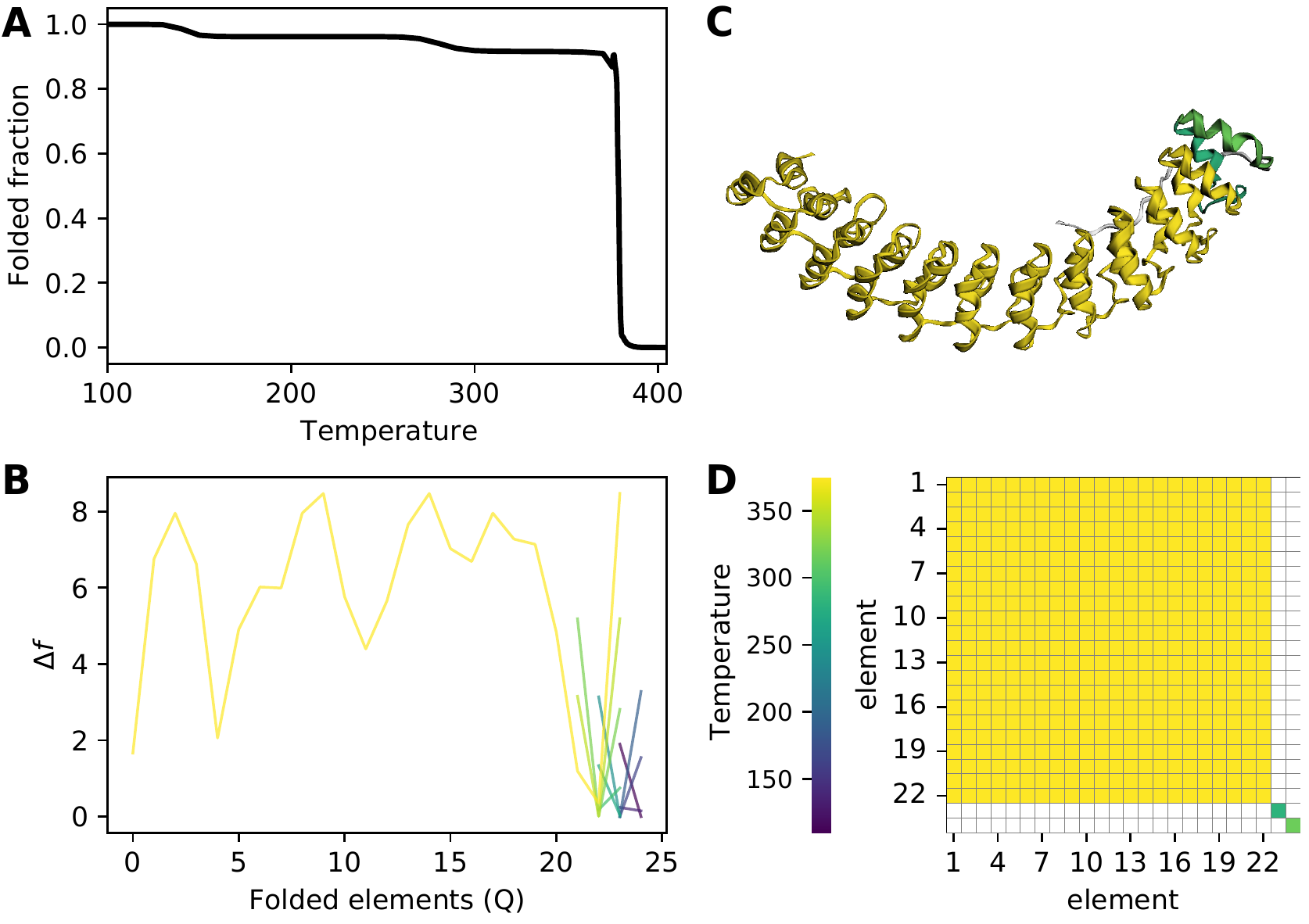}
\caption{\textbf{Simulation results for D34.} A. Thermal unfolding curve. B. Approximate free energy profiles, coloured by temperature (same of C and D), with the number of folded elements $Q$ as reaction coordinate. C. PDB structure is coloured according to the folding temperature of each element. Purple fragments are the most unstable ones. D. Apparent domain matrix, coloured by domain folding temperatures.}
\label{fig:D34}
\end{figure}

We study the 24-element fragment of protein AnkyrinR known as D34. Fluorescence chemical denaturation curves suggest a folding via a stable intermediate. Two subdomains were characterised experimentally: a more cooperative N-terminal half of 12 elements and a less cooperative C-terminal half \cite{werbeck2007probing}. In our simulations, although approximate free energy profiles revealed high energy intermediates, the experimentally described domains are not found. Only the last two elements separate from a large 22-elements highly cooperative domain (Fig. \ref{fig:D34}).

\subsection{Supplementary methods}
\subsubsection{Sequence data}
The \textit{full dataset} was built scanning UniprotKB \cite{boutet2007uniprotkb} with a structurally-derived Hidden Markov Models for single internal, N-terminal or C-terminal repeats \cite{Parra2015} using the \texttt{hmmsearch} tool at default parameters on single sequence files. Detected elements were forced to be 33 residues long and aligned, accepting deletions as gap characters '-', eliminating rarely occurring \cite{galpern2020large} insertions and filling missing terminal residues with gaps. Consecutive repeats were concatenated into arrays, allowing more than one array per protein sequence if elements were more than 67 residues away from each other. More details about the full dataset can be found in a previous work \cite{galpern2020large}. Given that evolutionary model was trained only on internal repeats, we discard sequences with 1 or 2 repeats. For minimizing phylogenetic bias, we cluster by full sequence similarity using CD-hit \cite{cdhit} at 90\% cutoff and maximum length difference of 32 residues, keeping randomly a single sequence per cluster. We label as \textit{natural dataset} the corresponding 117304 arrays and 740458 repeats on 109390 protein sequences. In order to build the 4020 array \textit{selected set} with which we ran Ising model simulations, we eliminate arrays where insertions, deletions, unknown residues were detected from the \textit{natural dataset}. In addition, arrays in which at least one terminal repeat was truncated are discarded. Finally we eliminate short 3-repeat arrays, because terminal repeats cover $2/3$ of the array and also the 3 remaining long arrays with more than 36 repeats, for computation time reasons.




\subsubsection{Evolutionary energy and experimental free folding energy}
\begin{figure}
\centering
\includegraphics[width=1.\linewidth]{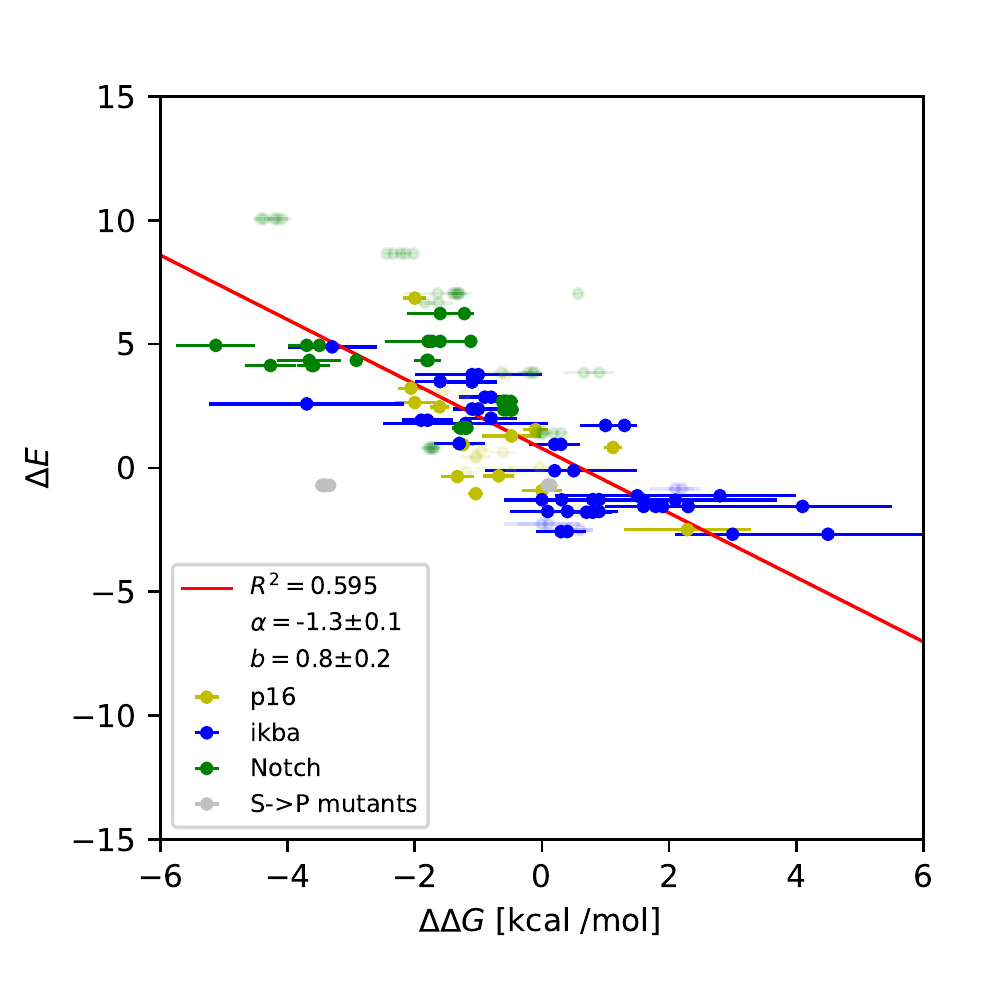}
\caption{Evolutionary energy change between a wildtype protein and a mutated variant as a function of the experimental unfolding free energy $\Delta \Delta G$. Error bars indicate the experimental standard deviation. Serine to Proline mutants (grey) have been described as structurally disruptive, hence they were not used for fitting.}
\label{fig:DDG}
\end{figure}

We considered the stability change by point mutations $\Delta \Delta G$ for three different ANK proteins: p16\textsuperscript{INK4A} \cite{p16_guo2010contributions,p16_tang1999stability,p16_tang2003sequential}, Notch \cite{notch_street2005improved,notch_tripp2008rerouting} and I$\kappa$B$\alpha$ \cite{ikba_devries2011folding,ikba_ferreiro2007stabilizing}. To compare with evolutionary energy change, we discarded mutations in terminal repeats, because the evolutionary model was trained only in internal repeats and Serine to Proline mutations. 
We compared the experimental unfolding energy difference between mutants and wild type $\Delta \Delta G$ and we compute $\Delta E$ for the same mutants \ref{fig:DDG}. We made a Linear fit and used the slope $\alpha=-1.3\pm0.1$ to convert one energy into the other.

\subsubsection{Entropy determination}
\begin{figure}
\centering
\includegraphics[width=1.\linewidth]{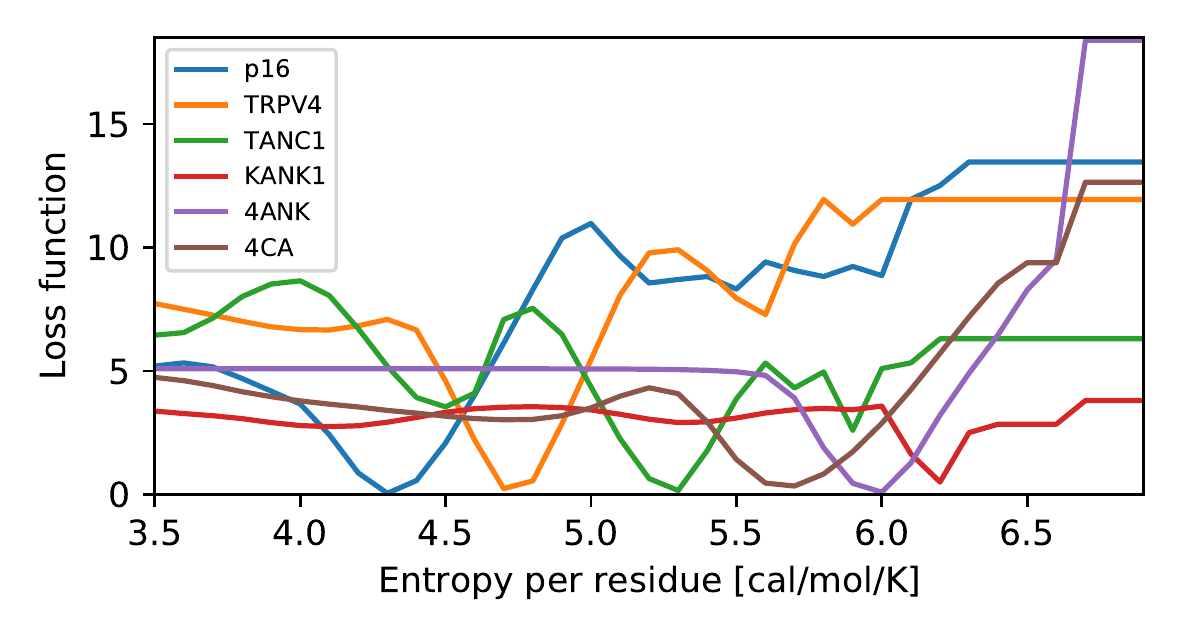}
\caption{We used least squares as loss function to get an optimal entropy per residue $s$ for the thermal unfolding curve of studied proteins.}
\label{fig:s_fit}
\end{figure}

\begin{figure*}
\centering
\includegraphics[width=1.0\linewidth]{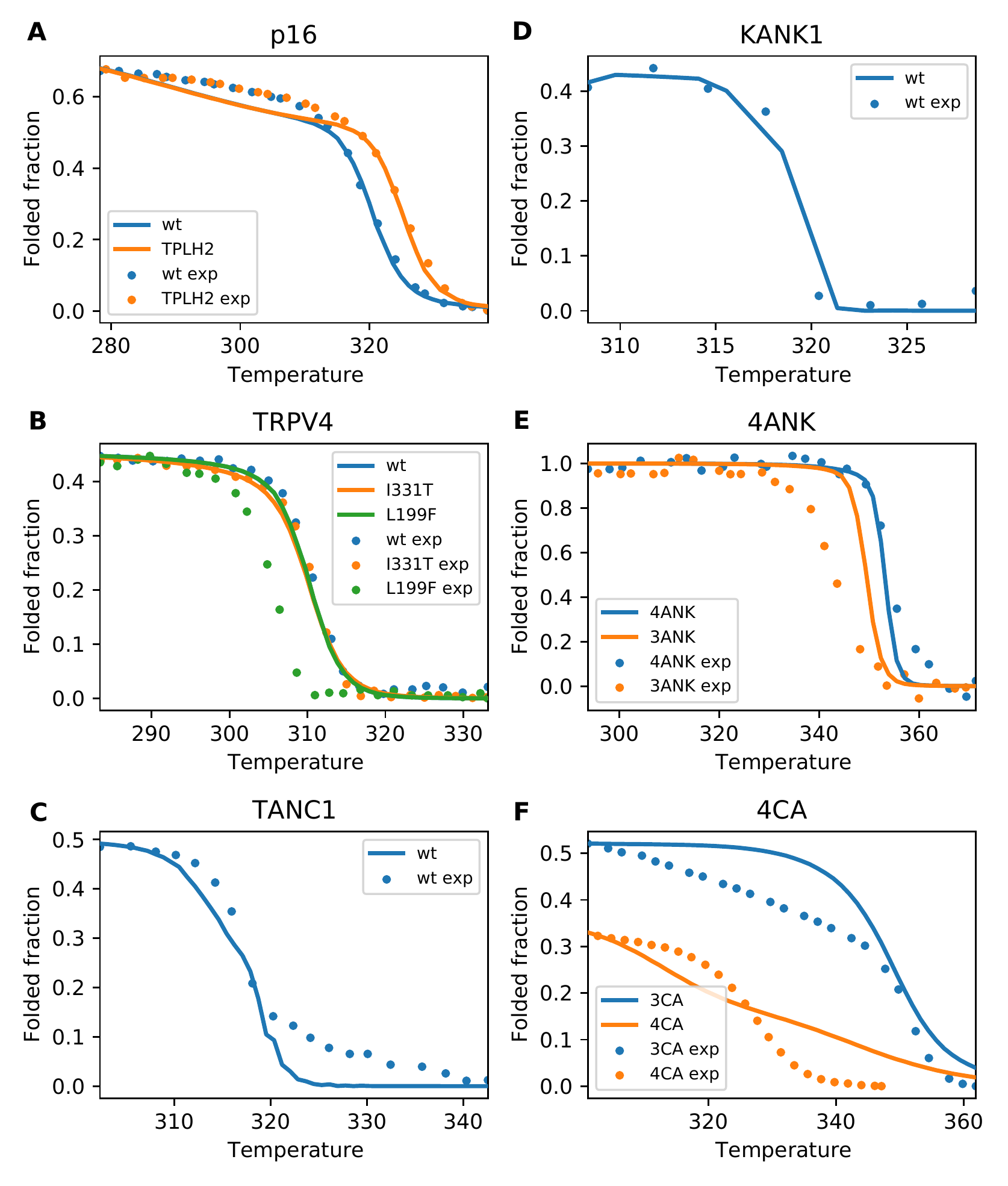}
\caption{\textbf{Optimal unfolding curves and mutation effect.} Simulations (lines) and rescaled CD signal to match the maximum folded fraction in the temperature range (points). Optimal $s$ found for each wild-type protein was used to simulate also mutated sequences if available. A. p16 wildtype and TPLH2 mutant \cite{p16_guo2010contributions}. B. TRPV4-ARD wildtype, I331T and L199F mutants \cite{trpv4_inada2012structural}. C TANC1-ARD \cite{tanc1_yang2019purification}. D. Kidney ANK 1 \cite{kank1_pan2018structural}. E. DARPins 4ANK ($s$ used) and 3ANK \cite{mosavi2002consensus}. F. DARPins 4CA ($s$ used) and 3CA \cite{schofield} }
\label{fig:mutants}
\end{figure*}

\begin{figure}
\centering
\includegraphics[width=1.\linewidth]{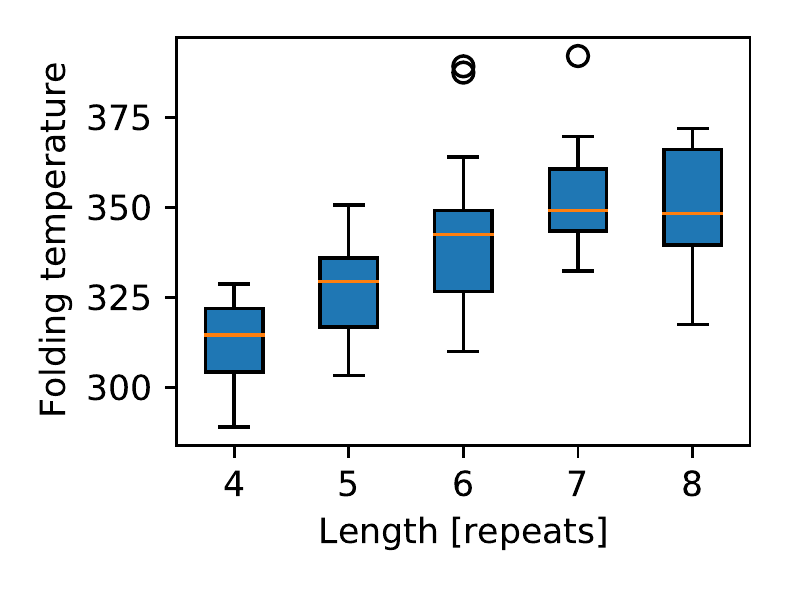}
\caption{Sigmoid fitted folding temperature for a library of $100$ DARPins as a function of sequence length. Stability increase with length was found.}
\label{fig:darpin}
\end{figure}

\begin{figure*}
\centering
\includegraphics[width=1.0\linewidth]{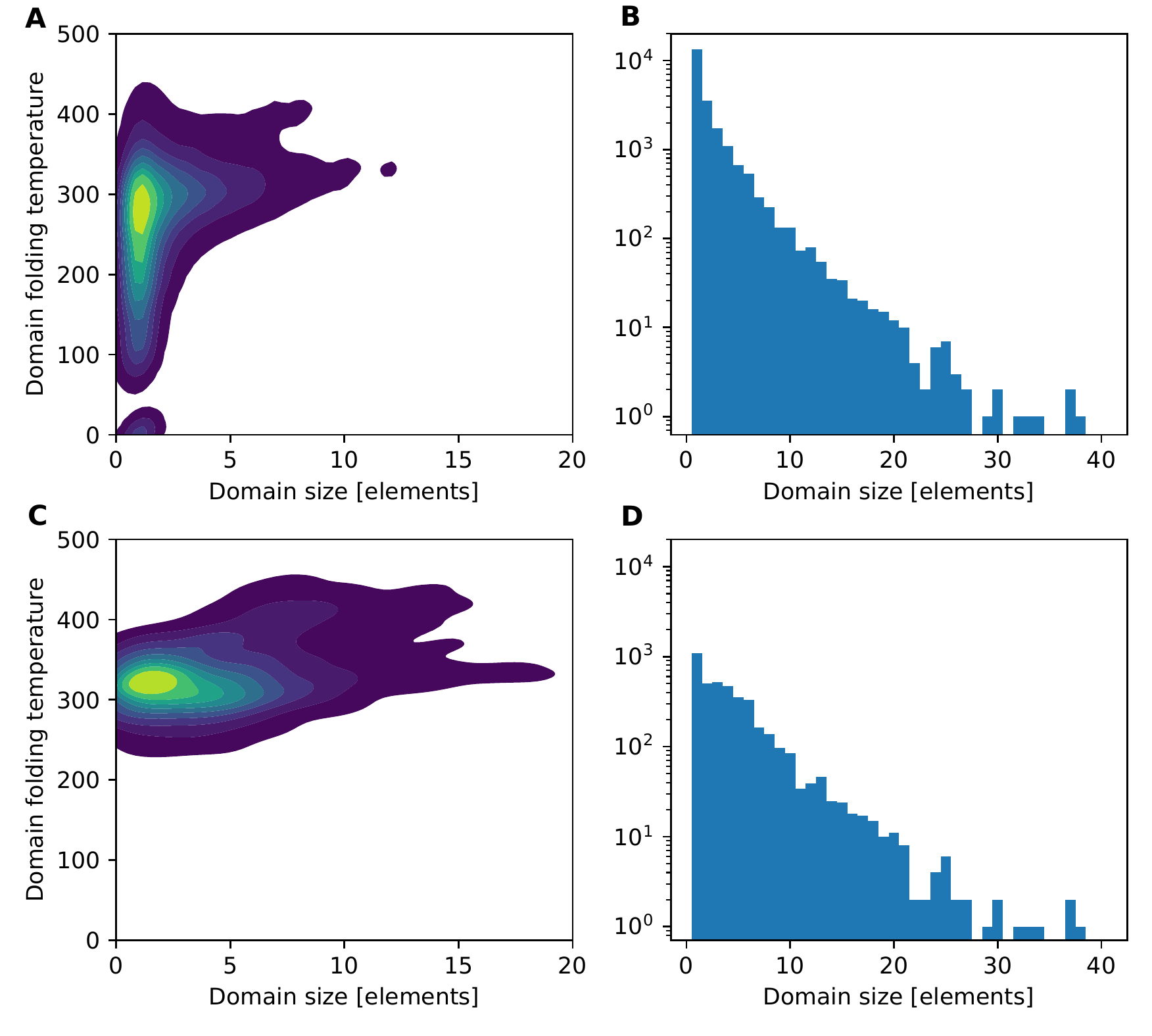}
\caption{A. Domain folding temperature and size heat-map. B Domain size distribution. Panels C and D show results only for the first domain to fold (nucleus) of each array.}
\label{fig:domains_sup}
\end{figure*}

\begin{figure}
\centering
\includegraphics[width=1.\linewidth]{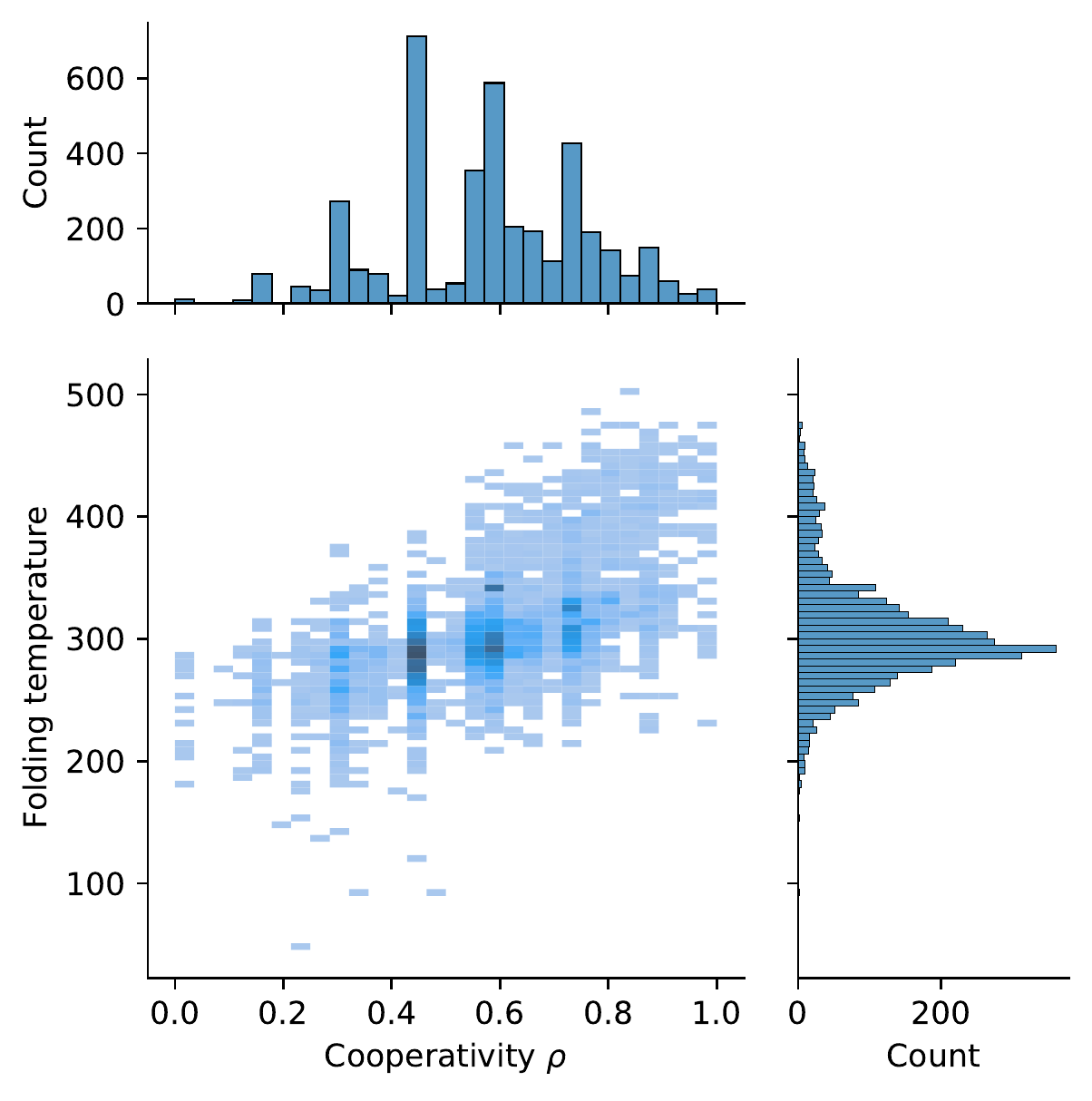}
\caption{Arrays folding temperature and cooperativity score $\rho$ joint histogram.}
\label{fig:rho_tf}
\end{figure}

\begin{figure}
\centering
\includegraphics[width=1.\linewidth]{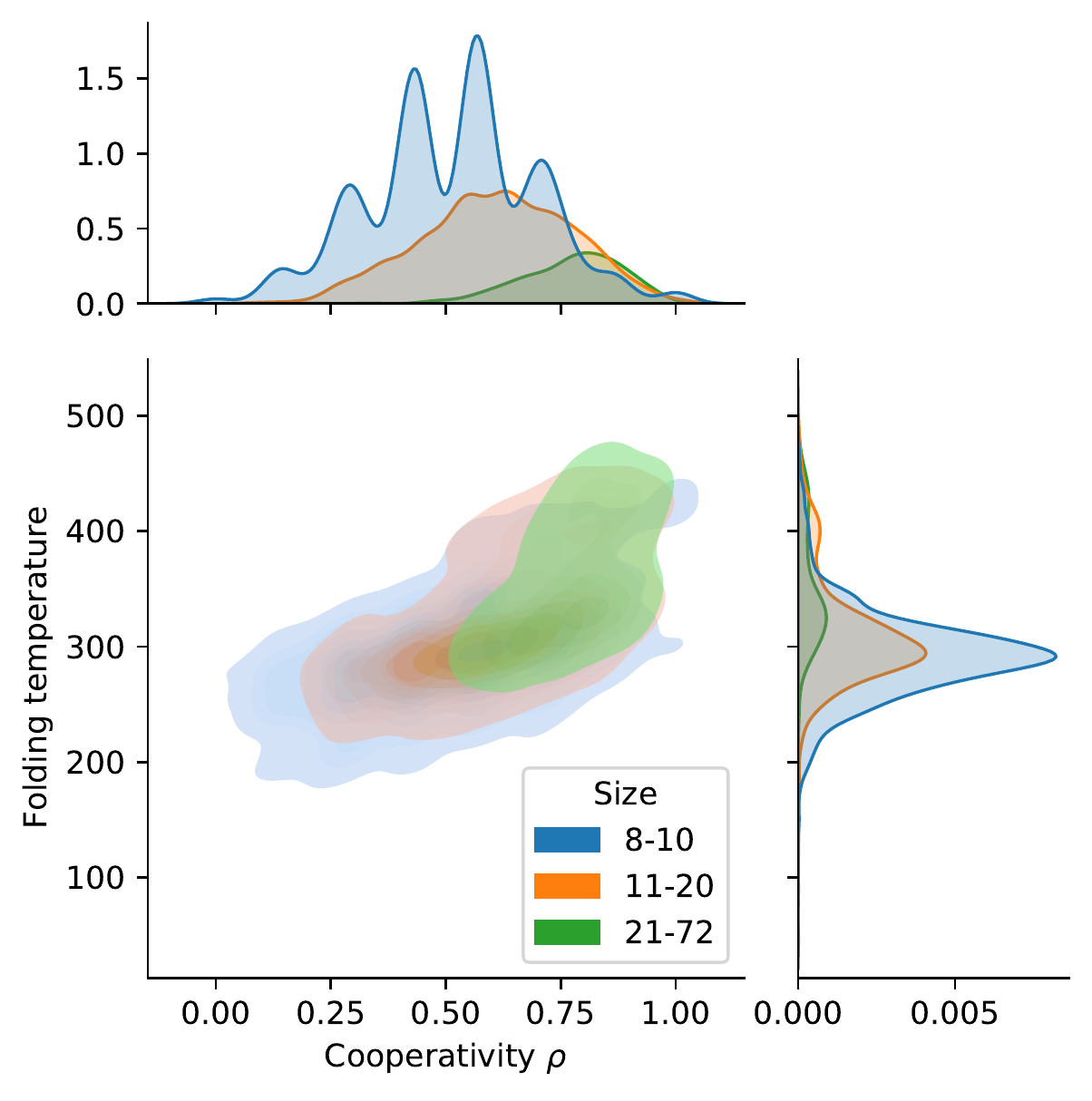}
\caption{Arrays folding temperature and cooperativity score $\rho$ joint distribution for short arrays (blue), intermediate (orange) and large ones (green).}
\label{fig:rho_tf_len}
\end{figure}

\begin{figure}
\centering
\includegraphics[width=1.\linewidth]{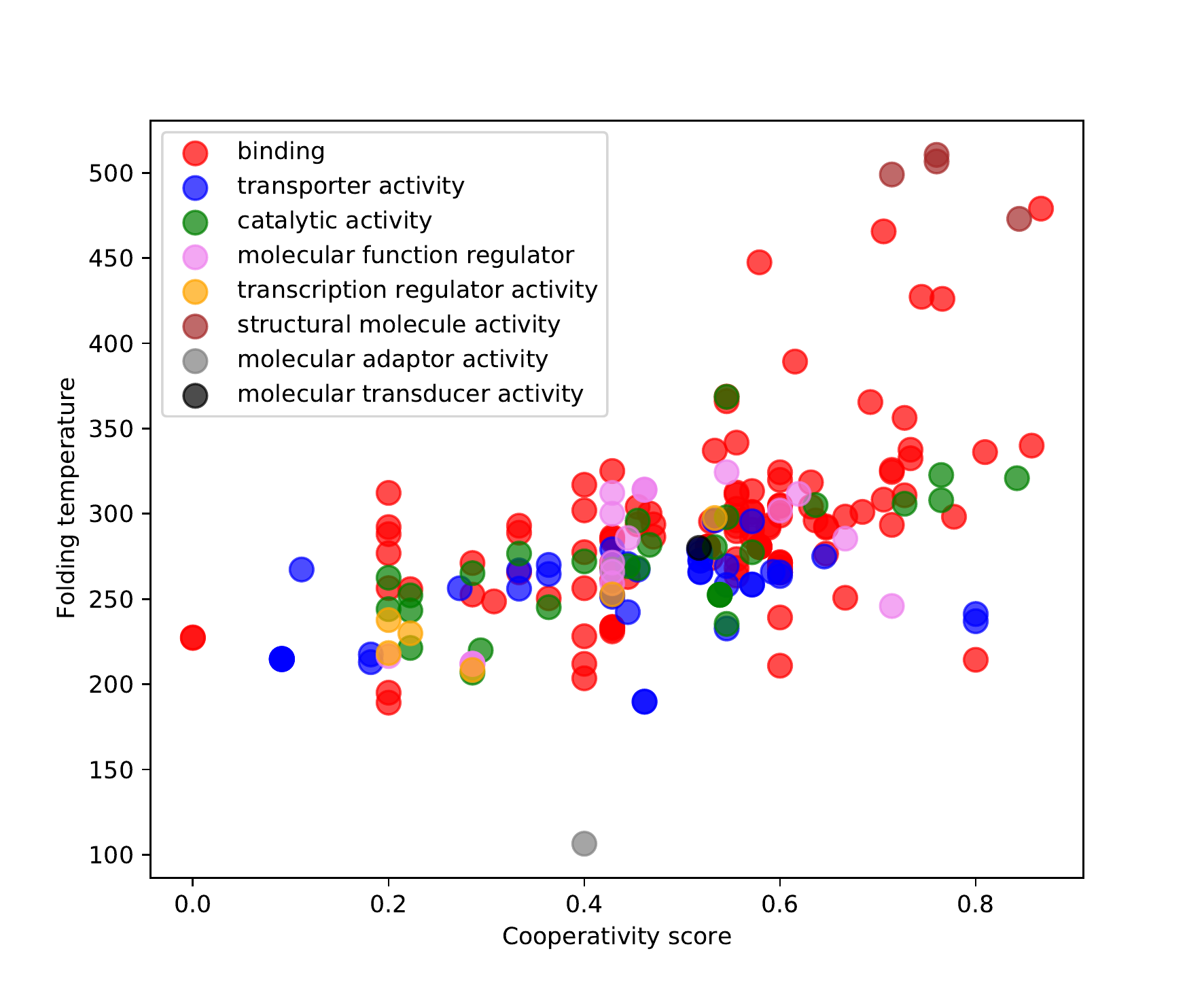}
\caption{Arrays with Gene Ontology annotations with experimental evidence of molecular function (colours) folding temperature (y-axis) and cooperativity score (x-axis).}
\label{fig:go}
\end{figure}

\begin{figure*}
\centering
\includegraphics[width=0.8\linewidth]{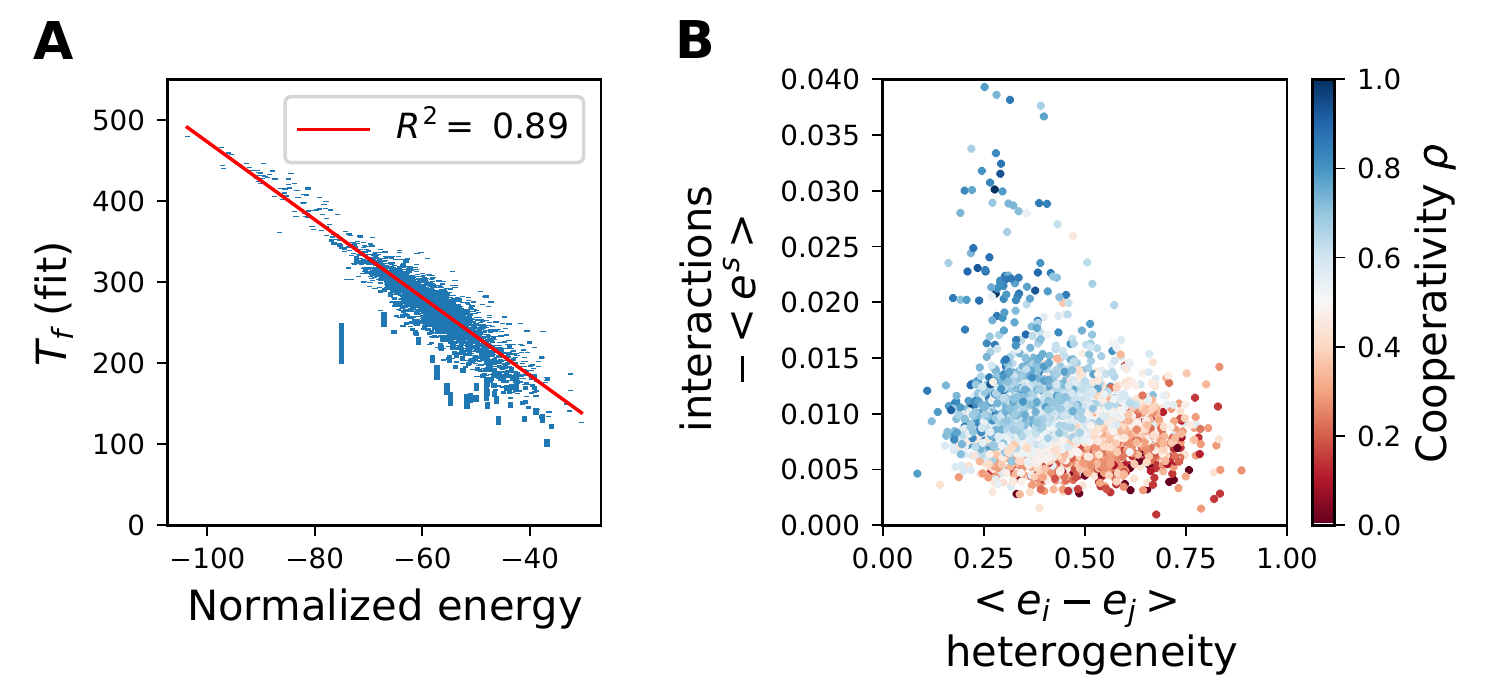}
\caption{\textbf{Generated sequence dataset results. }A. Array $T_f$ estimations with sigmoid function fits as a function of length-normalized co-evolutionary energy of the respective sequences. B. Cooperativity score $\rho$ is showed in a color scale on a plane defined by the average normalized internal energy difference $\langle e^i_j - e^i_k \rangle$ and the average non-zero normalized surface energies $- \langle e^s \rangle$.}
\label{fig:sup_fig5}
\end{figure*}

Optimal values for the entropy per residue $s$ were found by comparing simulated fraction folded curves with experimental ones. We used reversible thermal unfolding curves available in the literature for p16\textsuperscript{INK4A} wildtype and TPLH2 mutant \cite{p16_guo2010contributions}, TANC1 ANK repeat domain \textit{re+2m} mutant \cite{tanc1_yang2019purification}, TRPV4 ANK repeat Domain wildtype, I331T and L199F mutants \cite{tanc1_yang2019purification}, Kidney ANK repeat containing protein KANK1 \cite{kank1_pan2018structural}, DARPins 4ANK, 3ANK \cite{mosavi2002consensus}, 4CA and 3CA \cite{schofield}. Circular Dichroism (CD) at 222 nm experimental curves both raw and normalized as fraction folded were extracted from figures with the \textit{WebPlotDigitizer} web server \cite{webplotdigitalizer} point by point or if it was not possible using the auto-extracting tool. If there were more available, we selected 25 fixed uniformly distributed temperatures in the experimental range. Correspondent sequence alignments were curated by hand and are available upon request.
On wildtype sequences, we ran the model scanning entropy per residue $s$ at the selected temperatures. For each $s$ value, simulated fraction folded was normalized to reach the maximum experimental value at the lower temperature and we found the optimal entropy by least squares (Fig. \ref{fig:s_fit}). For each wildtype optimal $s$, we present fraction folded both for wildtype and mutants in fig. \ref{fig:mutants}.

\subsubsection{Gene Ontology analysis}
We checked for Gene Ontology (GO) \cite{gene2021gene} molecular function annotations for all the UnirprotKB \cite{boutet2007uniprotkb} (September 2021) arrays of the \textit{full dataset}. We considered that proteins achieve functions globally, hence we linked all the GO entries of the corresponding protein with the arrays detected on that sequence. We only worked experimental evidence codes and we kept one array per sequence, reducing the set to 217 proteins on which we ran the folding Ising model. In order to divide the data into comparable classes, we used GOATOOLS python library \cite{goatools} to get the depth-1 class for all the annotations.

\bibliographystyle{pnas}

\end{document}